\documentclass[notitlepage,aps,prl,twocolumn,superscriptaddress,amsmath,showpacs,tightenlines,longbibliography,10pt]{revtex4-1}
\usepackage{amssymb}
\usepackage{amsmath}
\usepackage{dcolumn}
\usepackage{graphicx}
\usepackage{mathrsfs}
\usepackage{subfigure}
\usepackage{booktabs}
\usepackage{color}
\usepackage{docmute}
\usepackage{hyphenat}
\usepackage{CJKutf8}
\usepackage{multirow}
\usepackage{enumerate}

\newdimen\origiwspc
\newdimen\origiwstr

\newcommand{\average}[1]{\mbox{$\langle#1\rangle$}}

\makeatletter
\newcommand{\raisemath}[1]{\mathpalette{\raisem@th{#1}}}
\newcommand{\raisem@th}[3]{\raisebox{#1}{$#2#3$}}
\makeatother

\usepackage{url}
\usepackage[colorlinks]{hyperref}
\hypersetup{%
	plainpages=true,
	breaklinks=true,
	hypertexnames=false,
	pageanchor=true,
	bookmarksnumbered=true,
	colorlinks=true,
	linkcolor={blue},
	citecolor={red},
	urlcolor={blue},
	anchorcolor={black}
}

\begin{document}
	
\title{Exponentially Improved Dispersive Qubit Readout with Squeezed Light}

\affiliation{Center for Joint Quantum Studies and Department of Physics, School of Science, Tianjin University, Tianjin 300350, China\\
	$^2$Theoretical Quantum Physics Laboratory, Cluster
	for Pioneering Research, RIKEN, Wako-shi, Saitama 351-0198, Japan\\
	$^3$Institute of Spintronics and Quantum Information, Faculty of Physics, Adam Mickiewicz University,
	61-614 Pozna\'n, Poland\\
	$^4$Center for Quantum Computing, RIKEN, Wako-shi, Saitama 351-0198, Japan\\
	$^5$Department of Physics, The University of Michigan,
	Ann Arbor, Michigan 48109-1040, USA\\
	$^6$Tianjin Key Laboratory of Low Dimensional Materials Physics and Preparing Technology, Tianjin University, Tianjin 300350, China}

\author{Wei Qin$^{1,2,6,}$}
\email{qin.wei@tju.edu.cn}

\author{Adam Miranowicz$^{2,3}$}

\author{Franco Nori$^{2,4,5}$}

\begin{abstract}
It has been a long-standing goal to improve dispersive qubit readout with squeezed light. However, injected external squeezing (IES) {\it cannot} enable a practically interesting increase in the signal-to-noise ratio (SNR), and simultaneously, the increase of the SNR due to the use of intracavity squeezing (ICS) is even {\it negligible}. Here, we {\it counterintuitively} demonstrate that using IES and ICS together can lead to an {\it exponential} improvement of the SNR for any measurement time, corresponding to a measurement error reduced typically by many orders of magnitude. More remarkably, we find that in a short-time measurement, the SNR is even improved exponentially with {\it twice} the squeezing parameter. As a result, we predict a fast and high-fidelity readout. This work offers a promising path toward exploring squeezed light for dispersive qubit readout, with immediate applications in quantum error correction and fault-tolerant quantum computation.
\end{abstract}

\date{\today}

\maketitle

\emph{Introduction.---}Squeezed light is a powerful resource in modern quantum technologies~\cite{scully1997book,drummond2004quantum,AgarwalBook2013}. It has been widely used for various applications, including quantum key distribution~\cite{ralph1999continuous,cerf2001quantum,madsen2012continuous,peuntinger2014distribution}, mechanical cooling~\cite{asjad2016suppression,asjad2019optomechanical,clark2017sideband,lau2020ground}, light-matter interaction enhancement~\cite{lu2015squeezed,qin2018exponentially,leroux2018enhancing,ge2019trapped,burd2021quantum,villiers2024dynamically,qin2024quantum}, and even quantum advantage demonstration~\cite{zhong2020quantum,zhong2021phase,madsen2022quantum}. In particular, such nonclassical light plays a central role in high-precision quantum measurements~\cite{schnabel2017squeezed,lawrie2019quantum}, e.g., gravitational-wave detection~\cite{rabl2010quantum,aasi2013enhanced,grote2013first}, optomechanical motion sensing~\cite{iwasawa2013quantum,peano2015intracavity}, and longitudinal qubit readout~\cite{didier2015fast,eddins2018stroboscopic}. Despite these achievements, how to utilize squeezed light to improve dispersive qubit readout (DQR) still remains an unresolved challenge~\cite{barzanjeh2014dispersive,didier2015heisenberg,govia2017enhanced,liu2022noise,blais2021circuit}.

\begin{figure}[t]
	\centering
	\includegraphics[width=8cm]{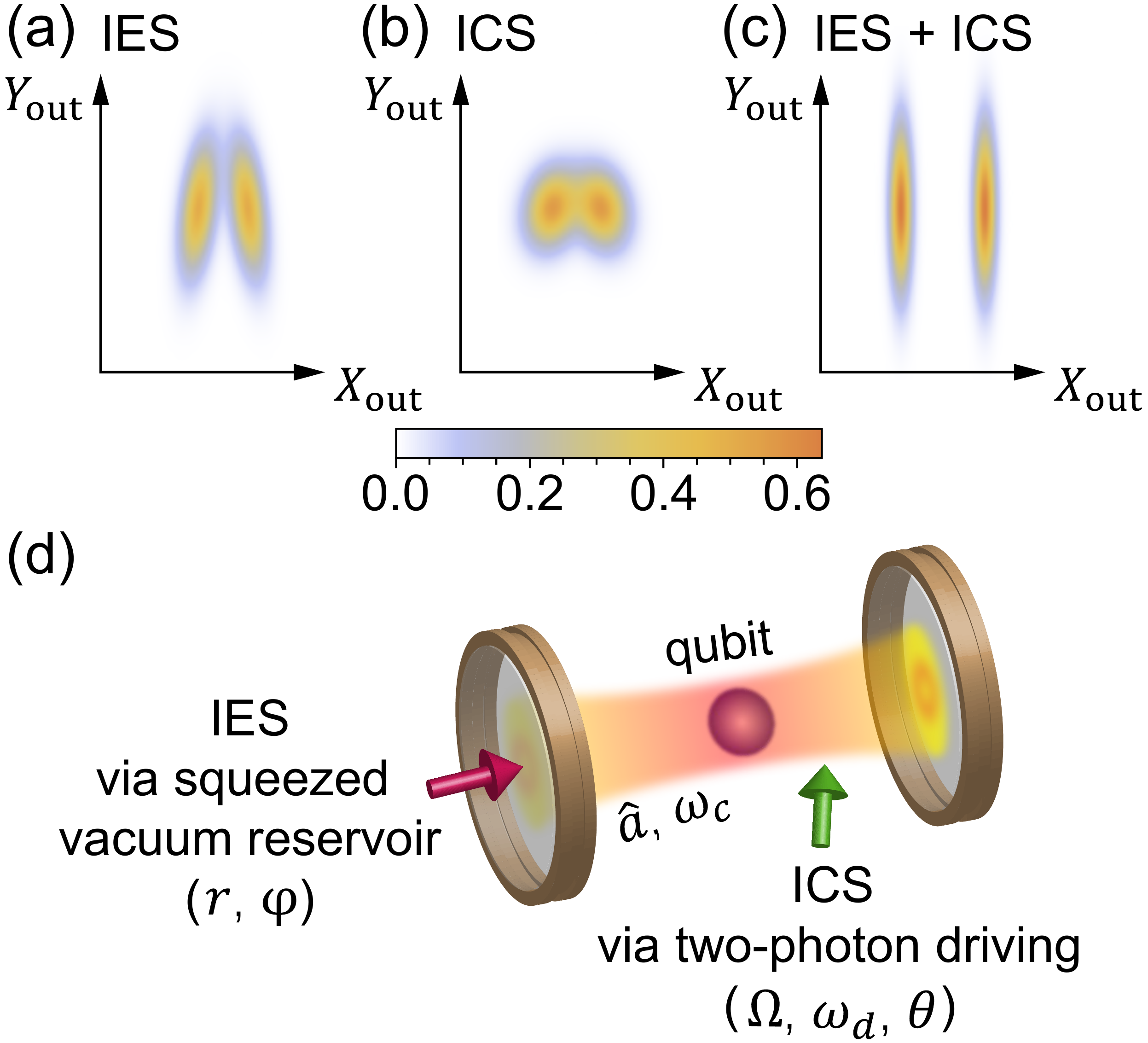}
	\caption{(a)-(c) Phase-space representation of DQR with IES, ICS, and these two simultaneously. The separate use of IES and ICS cannot enable a significant improvement of practical interest in the SNR, but their simultaneous use can. (d) Schematic of DQR with both IES and ICS. The qubit is dispersively coupled to the cavity mode $\hat{a}$ of frequency $\omega_{c}$. A squeezed vacuum reservoir (squeezing parameter $r$, reference phase $\varphi$) provides IES for the cavity, while a two-photon driving (amplitude $\Omega$, frequency $\omega_{d}$, phase $\theta$) is used to generate ICS.}\label{fig_schematics}
\end{figure}

DQR~\cite{wallraff2005approaching,krantz2016single,walter2017rapid,wang2019ideal,crippa2019gate,dassonneville2020fast}, as a common nondemolition readout, forms a crucial component of quantum error correction~\cite{schindler2011experimental,kelly2015state,krinner2022realizing} and fault-tolerant quantum computation~\cite{raussendorf2007fault,gambetta2017building}. In the readout, a qubit to be measured is dispersively coupled to a cavity working as the pointer, so that a qubit-state-dependent cavity resonance shift is induced and then measured by homodyne detection. Usually, this readout is required to be fast and of high fidelity, and exploiting squeezed states of light to improve such a readout is, therefore, highly desirable~\cite{blais2021circuit}. However, it has already been shown that injected external squeezing (IES) {\it cannot} significantly improve the signal-to-noise ratio (SNR) in an experimentally feasible way~\cite{didier2015heisenberg,supplement,kam2024sub}. The reason is attributed to a qubit-state-dependent rotation of squeezing and, thus, an increase in the overlap of the two pointer states [see Fig.~\ref{fig_schematics}(a)]. Furthermore, one might suggest the use of intracavity squeezing (ICS) generated, e.g., by a two-photon driving. However, in this case, the resulting improvement in the SNR is even {\it negligible}~\cite{supplement}. In addition to a detrimental rotation of squeezing, as in the case of IES, the reason is also related to the degree of squeezing that increases gradually from zero with the measurement time, as a result even causing a larger overlap of the pointer states [see Fig.~\ref{fig_schematics}(b)]. Hence, one can conclude that standard DQR {\it cannot} benefit well from the {\it separate} use of IES and ICS.

In this manuscript, we {\it counterintuitively} demonstrate that when IES and ICS are applied simultaneously, the readout SNR can have an {\it exponential} improvement for any measurement time. In our approach, the qubit-state information is mapped onto a Bogoliubov mode of the cavity, rather than the bare cavity mode as usual. This ensures a strong and measurement-time-independent degree of squeezing, and also avoids the qubit-state-dependent rotation of squeezing. Thus in sharp contrast to the case of using IES or ICS alone, the overlap of the pointer states is exponentially decreased [see Fig.~\ref{fig_schematics}(c)]. Note that a heuristic approach that can exponentially improve the SNR of DQR has been previously proposed, based on a quantum-mechanics-free subsystem~\cite{didier2015heisenberg}. But it needs to inject two-mode squeezed light into two coupled cavities, and measure a pair of readout modes. In contrast, our approach relies on a single cavity and single-mode squeezed light, therefore more suitable for the standard readout. More surprisingly, we find that the resulting SNR can scale as $e^{2r}$ for short-time measurements, rather than $e^{r}$ as given in Ref.~\cite{didier2015heisenberg}, indicating a fast and high-fidelity readout. Here, $r$ refers to the squeezing parameter. Such a giant improvement arises due to two aspects, one from antisqueezed vacuum fluctuations, which amplify the qubit-cavity dispersive coupling and, thus, the signal separation (i.e., the pointer-state separation), and the other from squeezed vacuum fluctuations, which reduce the measurement noise.

\emph{Physical model.---}The key idea underlying our proposal is shown in Fig.~\ref{fig_schematics}(d). The qubit is coupled to the cavity via a detuned interaction $\hat{H}_{\rm int}=g\left(\hat{a}^{\dagger}\hat{\sigma}_{-}+\hat{a}\hat{\sigma}_{+}\right)$, with a strength $g$. Here, $\hat{a}$ ($\hat{a}^{\dagger}$) is the annihilation (creation) operator of the cavity, and $\hat{\sigma}_{-}$ ($\hat{\sigma}_{+}$) is the lowering (raising) operator of the qubit. We assume that a squeezed vacuum reservoir, with a squeezing parameter $r$ and a reference phase $\varphi$, is injected into the cavity as IES~\cite{murch2013reduction,clark2017sideband,vahlbruch2018laser,malnou2019squeezed,xia2023entanglement}. To generate ICS, the cavity is further assumed to be pumped by a two-photon driving of amplitude $\Omega$, frequency $\omega_{d}$, and phase $\theta$. The system Hamiltonian in a frame rotating at $\omega_{d}$ is
\begin{equation}
\hat{H}=\Delta_{c}\hat{a}^{\dagger}\hat{a}+\frac{1}{2}\Delta_{q}\hat{\sigma}_{z}+\hat{H}_{\rm int}+\Omega\left(e^{i\theta}\hat{a}^{\dagger2}+e^{-i\theta}\hat{a}^{2}\right),
\end{equation}
where $\hat{\sigma}_{z}$ is a qubit Pauli operator, $\Delta_{c}=\omega_{c}-\omega_d/2$, and $\Delta_{q}=\omega_{q}-\omega_d/2$. Here, $\omega_{c}$ is the cavity frequency, and $\omega_{q}$ is the qubit transition frequency. The Langevin equation of motion for the cavity mode $\hat{a}$ is, therefore, given by
\begin{equation}\label{eq-motion-a}
\dot{\hat{a}}(t)=-i(\Delta_{c}-i\kappa/2)\hat{a}-i2\Omega e^{i\theta}\hat{a}^{\dagger}-ig\hat{\sigma}_{-}-\sqrt{\kappa}\hat{a}_{{\rm in}}(t),
\end{equation}
where $\kappa$ is the cavity-photon loss rate, and $\hat{a}_{{\rm in}}(t)$ represents the cavity input field. The correlations for the input-noise operator $\hat{\mathcal{A}}_{\rm in}(t)=\hat{a}_{{\rm in}}(t)-\average{\hat{a}_{{\rm in}}(t)}$ are $\average{\hat{\mathcal{A}}_{{\rm in}}(t)\hat{\mathcal{A}}_{{\rm in}}^{\dagger}(t^{\prime})}=\cosh^{2}(r)\delta(t-t^{\prime})$ and $\average{\hat{\mathcal{A}}_{{\rm in}}(t)\hat{\mathcal{A}}_{{\rm in}}(t^{\prime})}=\frac{1}{2}e^{i\varphi}\sinh(2r)\delta(t-t^{\prime})$, due to IES.

We below consider the case of $\Delta_{c}\neq0$, and introduce a Bogoliubov mode, $\hat{\beta}=\cosh(r_c)\hat{a}+e^{i\theta}\sinh(r_c)\hat{a}^{\dagger}$. Here, $\tanh(2r_c)=2\Omega/\Delta_{c}$. According to Eq.~(\ref{eq-motion-a}), the evolution of the mode $\hat{\beta}$ follows
\begin{equation}
\dot{\hat{\beta}}(t)=-i(\omega_{\rm sq}-i\kappa/2)\hat{\beta}-ig\hat{S}_{-}-\sqrt{\kappa}\hat{\beta}_{{\rm in}}(t),
\end{equation}
where $\omega_{\rm sq}=\sqrt{\Delta_{c}^{2}-4\Omega^{2}}$ is the resonance frequency of the mode $\hat{\beta}$, $\hat{S}_{-}=\cosh(r_c)\hat{\sigma}_{-}-e^{i\theta}\sinh(r_c)\hat{\sigma}_{+}$, and $\hat{\beta}_{\rm in}(t)=\cosh(r_c)\hat{a}_{{\rm in}}(t)+e^{i\theta}\sinh(r_c)\hat{a}_{{\rm in}}^{\dagger}(t)$. Upon choosing $r_c=r$ and $\theta-\varphi=\pi$, the correlations for the noise operator $\hat{\mathcal{B}}_{\rm in}(t)=\hat{\beta}_{\rm in}(t)-\average{\hat{\beta}_{\rm in}(t)}$ are~\cite{supplement}
\begin{equation}\label{eq-beta-input-correlation}
\average{\hat{\mathcal{B}}_{\rm in}(t)\hat{\mathcal{B}}_{\rm in}^{\dagger}(t^{\prime})}=\delta(t-t^{\prime}),\quad 
\average{\hat{\mathcal{B}}_{\rm in}(t)\hat{\mathcal{B}}_{\rm in}(t^{\prime})}=0.
\end{equation}
Therefore, $\hat{\mathcal{B}}_{\rm in}(t)$ can be now thought of as the vacuum noise. Note that similar techniques of noise elimination have been used, e.g., to enhance light-matter interactions~\cite{bartkowiak2014quantum,lu2015squeezed,zeytinouglu2017engineering,leroux2018enhancing,qin2018exponentially,lemonde2016enhanced,li2020enhancing,zhong2022quantum}, prepare nonclassical states~\cite{chen2021shortcuts}, generate squeezed lasing~\cite{sanchez2021squeezed}, and induce optical nonreciprocity~\cite{tang2022quantum}. However, these studies are based on the amplified fluctuations in an antisqueezed quadrature. To improve DQR, here we also exploit the reduced fluctuations in a squeezed quadrature, and as an overall result, we achieve an improved SNR scaling as $e^{2r}$ in a short-time measurement (see below). 

\begin{figure*}[t]
	\centering
	\includegraphics[width=17.5cm]{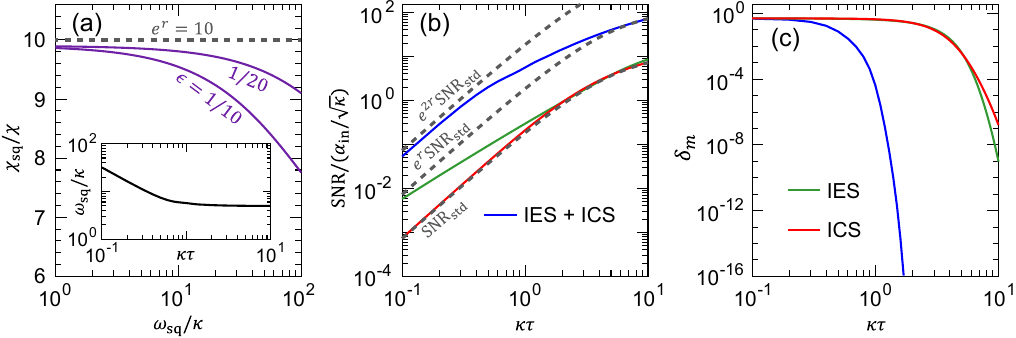}
	\caption{(a) Dispersive coupling enhancement, i.e, $\chi_{\rm sq}/\chi$, versus the effective cavity frequency $\omega_{\rm sq}$, for  $e^{r}=10$ and $\epsilon=1/10$, $1/20$. It is clearly seen that a significant and even an exponential enhancement can be obtained. Inset: $\omega_{\rm sq}$ required to achieve $|\average{\hat{M}}_{\uparrow}-\average{\hat{M}}_{\downarrow}|_{\perp}=0$ versus the measurement time $\kappa\tau$.  (b) SNR and (c) measurement error $\delta_m$ versus $\kappa\tau$. The three dashed curves in (b) represent ${\rm SNR}_{\rm std}$ (i.e., the SNR of the standard readout with no squeezing), $e^{r}{\rm SNR}_{\rm std}$, and $e^{2r}{\rm SNR}_{\rm std}$, respectively. In the cases of separately using IES (green) and ICS (red), the optimization has been made for the squeezing parameter $r$ in the range $1\leq e^{r} \leq10$, while we have set $\epsilon=1/20$ and $e^{r}=10$ in the readout of using both IES and ICS (blue). In (c), $\alpha_{\rm in}/\sqrt{\kappa}=1$, and in all plots, $\chi=\kappa/2$.}\label{fig_SNR}
\end{figure*}

Furthermore, we assume $(\Delta_{q}+\omega_{{\rm sq}})\gg g\cosh(r)$, and make a dispersive approximation~\cite{gamel2010time}. The effective dynamics of the system can be described by
\begin{equation}\label{eq-beta-evolution}
\dot{\hat{\beta}}(t)=-i(\omega_{\rm sq}+\chi_{\rm sq}\hat{\sigma}_{z}-i\kappa/2)\hat{\beta}-\sqrt{\kappa}\hat{\beta}_{{\rm in}}(t).
\end{equation}
Here, the $\hat{\sigma}_{z}$ term corresponds to a dispersive coupling, $\hat{V}_{\rm sq}=\chi_{\rm sq}\hat{\beta}^{\dagger}\hat{\beta}\hat{\sigma}_z$, of the qubit and the mode $\hat{\beta}$~\cite{supplement}, with a strength
\begin{equation}\label{eq-enhanced-dispersive-coupling}
	\chi_{\rm sq}=\chi\left\{\cosh(r)+\sinh^{2}(r)/[\cosh(r)+2\omega_{\rm sq}\epsilon/g]\right\},
\end{equation}
where $\chi=g\epsilon$ with $\epsilon= g\cosh(r)/(\Delta_{q}-\omega_{{\rm sq}})$. The dispersive coupling of the qubit and the bare mode $\hat{a}$, used for standard DQR, is usually obtained by applying a dispersive approximation to the detuned interaction $\hat{H}_{\rm int}$, and is given by $\hat{V}_{0}=\chi_{0}\hat{a}^{\dagger}\hat{a}\hat{\sigma}_{z}$, where $\chi_{0}=g \epsilon_{0}$ with $\epsilon_{0}=g/\left(\omega_q-\omega_c\right)$. It is seen that in our readout, the qubit-state information is mapped onto the Bogoliubov mode $\hat{a}_{\rm sq}$, rather than the bare mode $\hat{a}$ as in the standard readout; i.e., the qubit shifts the resonance of the mode $\hat{a}_{\rm sq}$, rather than the mode $\hat{a}$. 

Note that the parameters $\epsilon_{0}$ and $\epsilon$ determine the validity or accuracy of the dispersive approximations applied for $\hat{V}_{0}$ and $\hat{V}_{\rm sq}$, respectively. To compare $\hat{V}_{0}$ and $\hat{V}_{\rm sq}$ fairly, they need to have the same validity (i.e., $\epsilon_0=\epsilon$) and, in such a case, $\chi$ can be regarded as the dispersive coupling strength $\chi_{0}$ (i.e., $\chi=\chi_{0}$) (see~\cite{supplement} for more details). Consequently, as plotted in Fig.~\ref{fig_SNR}(a), $\chi_{\rm sq}$ becomes significantly enhanced by squeezing, compared to $\chi_0$ (i.e., $\chi$). Particularly, for $\epsilon\ll1$, we obtain an exponential enhancement, 
\begin{equation}\label{eq-exponentially-enhanced-chi}
\chi_{\rm sq}\simeq\chi \exp(r),
\end{equation}
which, physically, originates from the amplification of the qubit-cavity coupling from $g$ to $\simeq ge^{r}$ by the antisqueezing of vacuum fluctuations. As demonstrated below, such an enhancement can exponentially increase the signal separation and, thus, the readout SNR for short-time measurements.

\emph{Exponentially enhanced DQR.---}The output quadrature measured via homodyne detection is $\hat{\mathcal{Z}}_{{\rm out}}(t)=\hat{a}_{{\rm out}}(t)e^{-i\phi_{h}}+\hat{a}_{{\rm out}}^{\dagger}(t)e^{i\phi_{h}}$. Here, $\hat{a}_{{\rm out}}(t)=\cosh(r)\hat{\beta}_{{\rm out}}(t)-e^{i\theta}\sinh(r)\hat{\beta}_{{\rm out}}^{\dagger}(t)$, where $\hat{\beta}_{{\rm out}}(t)=\hat{\beta}_{{\rm in}}(t)+\sqrt{\kappa}\hat{\beta}$, represents the cavity output field, and $\phi_{h}$ is the measurement angle.
The SNR, an essential parameter to quantify homodyne detection, is defined as
\begin{equation}\label{eq-SNR-definition}
{\rm SNR}=|\average{\hat{M}}_{\uparrow}-\average{\hat{M}}_{\downarrow}|(\average{\hat{M}_{N}^{2}}_{\uparrow}+\average{\hat{M}_{N}^{2}}_{\downarrow})^{-1/2}.
\end{equation}
Here, $\hat{M}=\sqrt{\kappa}\int_{0}^{\tau}dt\hat{\mathcal{Z}}_{{\rm out}}(t)$ is the measurement operator with a measurement time $\tau$, $\hat{M}_{N}=\hat{M}-\average{\hat{M}}$ is the fluctuation-noise operator, and $\{\downarrow,\uparrow\}$ labels the qubit state. Note that although our proposal is based on the coupling $\hat{V}_{\rm sq}$ in the squeezed frame, the SNR in Eq.~(\ref{eq-SNR-definition}) is still given in the original lab frame as usual. Therefore, the SNR improvement mentioned below is measurable.

We now consider the measurement noise $\average{\hat{M}_{N}^{2}}$. The cavity output-noise operator, $\hat{\mathcal{A}}_{{\rm out}}(t)=\hat{a}_{{\rm out}}(t)-\average{\hat{a}_{{\rm out}}(t)}$, can be expressed as $\hat{\mathcal{A}}_{{\rm out}}(t)=\cosh(r)\hat{\mathcal{B}}_{{\rm out}}(t)-e^{i\theta}\sinh(r)\hat{\mathcal{B}}_{{\rm out}}^{\dagger}(t)$.
Here, $\hat{\mathcal{B}}_{\rm out}(t)=\hat{\beta}_{{\rm out}}(t)-\average{\hat{\beta}_{{\rm out}}(t)}$. From Eqs.~(\ref{eq-beta-input-correlation}, \ref{eq-beta-evolution}), we find 
$\average{\hat{\mathcal{B}}_{\rm out}(t)\hat{\mathcal{B}}_{\rm out}^{\dagger}(t^{\prime})}=\delta(t-t^{\prime})$ and $\average{\hat{\mathcal{B}}_{\rm out}(t)\hat{\mathcal{B}}_{\rm out}(t^{\prime})}=0$. It follows that 
\begin{align}
\average{\hat{\mathcal{A}}_{{\rm out}}(t)\hat{\mathcal{A}}_{{\rm out}}^{\dagger}(t^{\prime})}=\;&\cosh^{2}(r)\delta(t-t^{\prime}),\\
\average{\hat{\mathcal{A}}_{{\rm out}}(t)\hat{\mathcal{A}}_{{\rm out}}(t^{\prime})} =\;&-\frac{1}{2}e^{i\theta}\sinh(2r)\delta(t-t^{\prime}),
\end{align}
and then that 
\begin{equation}
\average{\hat{M}_{N}^{2}}=\kappa\tau\left[\cosh(2r)-\cos(2\phi_{h}-\theta)\sinh(2r)\right].
\end{equation}
Clearly, $\average{\hat{M}_{N}^{2}}$ is qubit-state independent for any measurement time, in stark contrast to the case of using IES or ICS alone [see Figs.~\ref{fig_schematics}(a, b)]. Furthermore, choosing $2\phi_{h}=\theta$ gives
\begin{equation}\label{eq-e-suppressed-noise}
\average{\hat{M}_{N}^{2}}=\kappa\tau \exp(-2r),
\end{equation}
indicating an exponential suppression of the measurement noise for any measurement time. 

Consider a coherent measurement tone, $\average{\hat{a}_{\rm in}(t)}=\alpha_{\rm in}e^{i\phi_{\rm in}}$. Since the signal separation is proportional to $|\average{\hat{\beta}_{\rm in}(t)}|$~\cite{supplement}, we therefore maximize $|\average{\hat{\beta}_{\rm in}(t)}|$ by assuming $2\phi_{\rm in}=\theta$, yielding $\average{\hat{\beta}_{\rm in}(t)}=\alpha_{\rm in}e^{r}e^{i\phi_{\rm in}}$. Note that in the optimal case of using IES or ICS alone, the signal separation perpendicular to the measurement direction, labelled $|\average{\hat{M}}_{\uparrow}-\average{\hat{M}}_{\downarrow}|_{\perp}$, always vanishes~\cite{supplement}, but not in the case of their simultaneous use. Thus for comparison, we require $|\average{\hat{M}}_{\uparrow}-\average{\hat{M}}_{\downarrow}|_{\perp}=0$ in our proposal, although reducing the SNR slightly. For a given measurement time, this requirement can be exactly satisfied by tuning $\omega_{\rm sq}$, as depicted in the inset of Fig.~\ref{fig_SNR}(a). 

Before presenting numerical simulations, let us first discuss the two limits, i.e., $\kappa\tau\rightarrow0$ and $\infty$. In the short-time limit $\kappa\tau\rightarrow0$, the SNR is given by~\cite{supplement}:
\begin{equation}\label{eq-SNR-short-time}
	{\rm SNR}\simeq0.81\exp(2r){\rm SNR}_{\rm std},
\end{equation}
where ${\rm SNR}_{{\rm std}}$ refers to the SNR of the standard readout with no squeezing. Surprisingly, we find that the SNR is improved exponentially with $2r$. There are two reasons for this. First, the measurement noise is exponentially reduced, as seen in Eq.~(\ref{eq-e-suppressed-noise}). Second, the signal separation, $\simeq0.27\alpha_{\rm in}\kappa^{-1/2}\tan(\psi_{\rm sq})\left(\kappa\tau\right)^{3}$, with $\tan(\psi_{\rm sq})=2\chi_{\rm sq}/\kappa$,
is increased by a factor $e^{r}$, due to the enhanced $\chi_{\rm sq}$ as given in Eq.~(\ref{eq-exponentially-enhanced-chi}). Instead, in the long-time limit $\kappa\tau\rightarrow\infty$, the signal separation, $\simeq4\alpha_{\rm in}\kappa^{-1/2}\sin(\psi_{\rm sq})\kappa\tau$, is not changed significantly by $\chi_{\rm sq}$ or $r$; thus, we also have an exponential improvement but with $r$, i.e.,
\begin{equation}\label{eq-SNR-long-time}
	{\rm SNR}\simeq\frac{\sin(\psi_{\rm sq})}{\sin(2\psi)}\exp(r)\,{\rm SNR}_{\rm std}.
\end{equation}
In Figs.~\ref{fig_SNR}(b, c), we plot the SNR and the measurement error, $\delta_m=1-\mathcal{F}_m$, for DQR using IES, ICS, and both of them. Here, $\mathcal{F}_m=\frac{1}{2}[1+{\rm erf}({\rm SNR}/2)]$ refers to the measurement fidelity. Note that in the case of ICS, we have defined  $r=\ln[(\kappa+4\Omega)/(\kappa-4\Omega)]$, which is the squeezing parameter of the cavity output field in the absence of the qubit. Clearly, for any measurement time, our approach can enable at least an exponential improvement with $r$, compared to the other approaches. Assuming realistic parameters of $\alpha_{\rm in}/\sqrt{\kappa}=1$, $\chi=\kappa/2$, and $\kappa=2\pi\times5$~MHz, a typical measurement time of $\tau=1/\kappa\simeq32$~ns results in ${\rm SNR}\simeq5.5$ for $e^{r}=10$, corresponding to a measurement error of $\delta_m\simeq4.4\times10^{-5}$. In stark contrast, we find ${\rm SNR}\simeq 0.18$, $0.29$, and $0.21$ for the readout with no squeezing, IES, and ICS, respectively. The corresponding measurement errors are  $\delta_m\simeq0.45$, $0.42$, and $0.44$, all approximately {\it six orders of magnitude} larger than what is obtained using our approach. Indeed, as analyzed in detail in Ref.~\cite{supplement}, the use of IES or ICS alone cannot enable a significant SNR increase of practical interest.  

\begin{figure}[t]
	\centering
	\includegraphics[width=8.56cm]{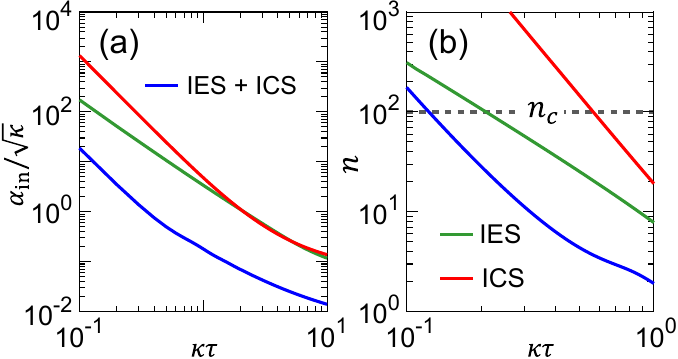}
	\caption{(a) Measurement-tone amplitude $\alpha_{\rm in}/\sqrt{\kappa}$ and (b) cavity-photon number $n$, required to achieve ${\rm SNR}=1$, versus $\kappa\tau$. In our approach, the photon number $n$ depends on the qubit state, and we depict the maximum values here. The horizontal dashed line indicates the critical photon number $n_c=100$. It is seen that our approach can keep $n$ well below $n_c$ at a much smaller $\kappa\tau$, compared to the other approaches. In both plots, the parameters are chosen as in Fig.~\ref{fig_SNR}(b).}\label{fig_n}
\end{figure}

\emph{Fast DQR.---}Standard DQR, although simple, cannot be improved arbitrarily by simply increasing the measurement-tone amplitude. The reason is because the cavity-photon number $n$ typically needs to be kept well below the critical photon number $n_c=1/4\epsilon^2$, to avoid the breakdown of the dispersive approximation~\cite{blais2004cavity}. This makes the readout very slow. However, as mentioned above, the SNR obtained using our approach is improved by a factor $e^{2r}$ in the short-time measurement, implying that a fast and high-fidelity readout can be achieved.

In Figs.~\ref{fig_n}(a, b), we depict the measurement-tone amplitude $\alpha_{\rm in}$ and the cavity-photon number $n$, needed to reach ${\rm SNR}=1$, for the three cases of using IES, ICS, and both of them. In our approach, the dispersive approximation is made for the mode $\hat{\beta}$, rather than the mode $\hat{a}$; therefore, the cavity-photon number, used to evaluate the validity of this approximation, is  $n=\average{\hat{\beta}^{\dagger}(t)\hat{\beta}(t)}$~\cite{supplement}. Clearly, our approach enables a much shorter measurement time. For example, we use a measurement tone of $\alpha_{\rm in}/\sqrt{\kappa}\simeq3.5$, corresponding to $n\simeq29$ cavity photons, to have ${\rm SNR}=1$ for a short measurement time of $\tau=0.2/\kappa\simeq6.4$~ns. Here, $\kappa=2\pi\times5$~MHz. However, to reach the same SNR at the same measurement time, the approaches based on IES and ICS need the much stronger measurement tones of $\alpha_{\rm in}/\sqrt{\kappa}\simeq52$ and $\simeq239$, respectively, resulting in $n\simeq107$ and $\simeq2238$ cavity photons, both higher than the critical photon number $n_c=100$. Note that, here, the standard readout of no squeezing has almost the same results as in the case of ICS. 

\begin{figure}[t]
	\centering
	\includegraphics[width=8.56cm]{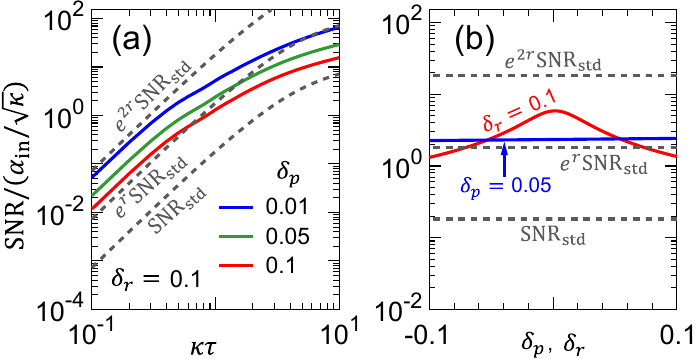}
	\caption{SNR versus $\kappa\tau$ for some fixed parameter mismatches in (a), and versus the parameter mismatch for $\kappa\tau=1$ in (b). We assumed $\delta_{p}=0.1$, $0.05$, $0.01$, and $\delta_{r}=0.1$ (see Ref.~\cite{supplement} for the $\delta_{r}=0.01$ case) in (a). In (b), the red and blue curves show the robustness against the squeezing-direction mismatch $\delta_{p}$ for $\delta_{r}=0.1$ and the squeezing-degree mismatch $\delta_{r}$ for $\delta_{p}=0.05$, respectively. In both plots, other parameters and what the three dashed curves represent are the same as in Fig.~\ref{fig_SNR}(b).}\label{fig_parameter_imperfect}
\end{figure}

\emph{Experimental feasibility.}---So far, we have discussed an ideal model where $r_c=r$ and $\theta-\varphi=\pi$. However, there are always some parameter mismatches in realistic experiments, such that $r_c=r+\delta_{r}$, and $\theta-\varphi=\pi+\delta_{p}$,
where $\delta_{r}$ and $\delta_{p}$ are the squeezing-degree and -direction mismatches, respectively. A detailed derivation of the SNR in such an imperfect case is given in Ref.~\cite{supplement}. We plot in Fig.~\ref{fig_parameter_imperfect} the SNR in the presence of these parameter mismatches. It is seen from Fig.~\ref{fig_parameter_imperfect}(a) that the exponential improvement in the SNR remains even in the presence of finite parameter mismatches. We also find from Fig.~\ref{fig_parameter_imperfect}(b) that the SNR is sensitive to the squeezing-direction mismatch but, interestingly, is very robust against the squeezing-degree mismatch. Clearly, an SNR improvement nearly exponential can still be achieved at a measurement time $\tau\sim1/\kappa$, even for large mismatches of $\delta_{p}=0.1$ and $\delta_{r}=0.1$; e.g., ${\rm SNR}\simeq0.72\exp(r){\rm SNR}_{\rm std}$, at $\tau=1/\kappa$. Moreover, our readout proposal is valid for a wide range of physical systems. Particularly, in superconducting quantum circuits, IES and ICS can be implemented using Josephson parametric amplifiers~\cite{yamamoto2008flux,clark2017sideband,murch2013reduction}. Hence, our proposal is experimentally feasible.

\emph{Conclusions.---}We have presented a method of simultaneously using IES and ICS to improve DQR. This method can enable at least an exponential improvement of the SNR with the squeezing parameter. In particular, the short-time SNR is improved exponentially with twice the squeezing parameter, therefore leading to a fast and high-fidelity readout. In stark contrast, using IES or ICS alone cannot make a significant and practically useful increase in the SNR. Our proposal opens a promising perspective for the use of squeezing, and could further stimulate more applications of squeezing for modern quantum technologies.

\begin{acknowledgments}
	W.Q.~acknowledges support of the National Natural Science Foundation of China (NSFC) via Grant No.~0401260012. A.M.~is supported by the Polish National Science Centre (NCN) under the Maestro Grant No.~DEC-2019/34/A/ST2/00081. F.N. is supported in part by:	
	Nippon Telegraph and Telephone Corporation (NTT) Research, the Japan Science and Technology Agency (JST)	[via the CREST Quantum Frontiers program Grant No. JPMJCR24I2,	the Quantum Leap Flagship Program (Q-LEAP), and the Moonshot R\&D Grant Number JPMJMS2061], and the Office of Naval Research (ONR) Global (via Grant No. N62909-23-1-2074).
\end{acknowledgments}

\emph{Note added.---}While completing this manuscript, we became aware of a very recent preprint~\cite{kam2024fast}, which also discusses the simultaneous use of IES and ICS for DQR. However, that work does not take full advantage of squeezing, and shows a completely different result, i.e., a modest (not an exponential as predicted in our present work) improvement in the SNR.


%

\end{document}


\title{Supplemental Material to:\\
	``Exponentially Improved Dispersive Qubit Readout with Squeezed Light"}

\affiliation{Center for Joint Quantum Studies and Department of Physics, School of Science, Tianjin University, Tianjin 300350, China\\
	$^2$Theoretical Quantum Physics Laboratory, Cluster
	for Pioneering Research, RIKEN, Wako-shi, Saitama 351-0198, Japan\\
	$^3$Institute of Spintronics and Quantum Information, Faculty of Physics, Adam Mickiewicz University,
	61-614 Pozna\'n, Poland\\
	$^4$Center for Quantum Computing, RIKEN, Wako-shi, Saitama 351-0198, Japan\\
	$^5$Department of Physics, The University of Michigan,
	Ann Arbor, Michigan 48109-1040, USA\\
	$^6$Tianjin Key Laboratory of Low Dimensional Materials Physics and Preparing Technology, Tianjin University, Tianjin 300350, China}

\author{Wei Qin$^{1,2,6,}$}
\email{qin.wei@tju.edu.cn}

\author{Adam Miranowicz$^{2,3}$}

\author{Franco Nori$^{2,4,5}$}

\makeatletter
\def\@hangfrom@section#1#2#3{\@hangfrom{#1#2#3}}
\makeatother

\maketitle

\section*{I\lowercase{ntroduction}}

\begin{quote}
	In this Supplemental Material, we investigate in detail dispersive qubit readout (DQR) with injected external squeezing (IES) in Sec.~\ref{Dispersive qubit readout only with injected squeezing}, and then the case with intracavity squeezing (ICS) in Sec.~\ref{Dispersive qubit readout only with intracavity squeezing}. We demonstrate that these two cases {\it cannot enable a significant and practically useful increase} in the signal-to-noise ratio (SNR). In Sec.~\ref{Enhanced dispersive coupling}, we show a detailed analysis of the qubit-cavity dispersive coupling enhanced by squeezing. In Sec.~\ref{Enhance dispersive qubit readout}, we present more details of the derivation of the SNR of DQR with both IES and ICS. In this case, we find that, in sharp contrast to the case of using IES or ICS alone, their simultaneous use can lead to {\it an exponential improvement} of the SNR. In particular, for a short time measurement, the SNR is improved {\it exponentially with twice the squeezing parameter}. Finally, the effects of parameter mismatches in realistic experiments on our readout proposal are discussed in Sec.~\ref{Effects of parameter mismatches on the readout}.
\end{quote}

\section{Dispersive qubit readout with injected external squeezing}
\label{Dispersive qubit readout only with injected squeezing}
In this section, we discuss dispersive qubit readout (DQR) with injected external squeezing (IES). Specifically, we derive in detail the measurement signal, the measurement noise, and as a result, the signal-to-noise ratio (SNR). We demonstrate that IES is able to exponentially improve the SNR only in the two impractical limits $\kappa\tau\rightarrow0$ and $\infty$, corresponding to a strong measurement tone and a long measurement time, respectively. However, in the regime $\kappa\tau\sim1$, which is of most interest in experiments, a qubit-state-dependent rotation of squeezing becomes dominant and increases the overlap of the pointer states, thus largely limiting the SNR improvement. {\it Thus, IES cannot significantly improve the SNR at an experimentally feasible measurement time.}

We begin with the readout Hamiltonian given by
\begin{equation}
\hat{H}=\chi\hat{a}^{\dagger}\hat{a}\hat{\sigma}_{z},
\end{equation}
where $\chi=g^{2}/\Delta$, with $g$ denoting the coupling of the qubit to the cavity and $\Delta$ denoting their detuning. Moreover, $\hat{\sigma}_{z}$ is a Pauli operator of the qubit. 
Correspondingly, the Langevin equation of motion for the cavity mode $\hat{a}$ reads
\begin{equation}\label{seq-motion-equation-a-injecting-squeezing}
\dot{\hat{a}}=-(\sigma\chi-i\kappa/2)\hat{a}-\sqrt{\kappa}\,\hat{a}_{\rm in}(t),
\end{equation}
where $\kappa$ is the photon loss rate of the cavity. Here, the qubit has been assumed to be in a definite state, such that the operator $\hat{\sigma}_{z}$ has been rewritten as a c-number $\sigma=\pm1$, corresponding to the excited and ground states of the qubit, respectively. Moreover, $\hat{a}_{\rm in}\left(t\right)$ represents the input field of the cavity. We assume that a squeezed vacuum reservoir, acting as IES, is injected into the cavity. In this case, the correlations for the noise operator $\hat{\mathcal{A}}_{\rm in}\left(t\right)=\hat{a}_{\rm in}\left(t\right)-\average{\hat{a}_{\rm in}\left(t\right)}$ are:
\begin{align}
\label{sq-correlations-injected-squeezing-01}
\average{\hat{\mathcal{A}}_{\rm in}^{\dagger}\left(t\right)\hat{\mathcal{A}}_{\rm in}\left(t^{\prime}\right)}=\;&\sinh^2(r)\delta(t-t^{\prime}), \quad 
\average{\hat{\mathcal{A}}_{\rm in}\left(t\right)\hat{\mathcal{A}}_{\rm in}^{\dagger}\left(t^{\prime}\right)}=\cosh^2(r)\delta(t-t^{\prime}),\\
\label{sq-correlations-injected-squeezing-02}
\average{\hat{\mathcal{A}}_{\rm in}\left(t\right)\hat{\mathcal{A}}_{\rm in}\left(t^{\prime}\right)}=\;&\frac{1}{2}e^{i\varphi}\sinh(2r)\delta(t-t^{\prime}),\quad
\average{\hat{\mathcal{A}}_{\rm in}^{\dagger}\left(t\right)\hat{\mathcal{A}}_{\rm in}^{\dagger}\left(t^{\prime}\right)}=\frac{1}{2}e^{-i\varphi}\sinh(2r)\delta(t-t^{\prime}),
\end{align}
where $r$ is the squeezing parameter of IES and $\varphi$ is the reference phase. It follows, after formally integrating Eq.~(\ref{seq-motion-equation-a-injecting-squeezing}), that 
\begin{equation}\label{seq-cavity-field-operator-injecting-squeezing}
\hat{a}(t)=\exp[-i(\sigma\chi-i\kappa/2)t]\hat{a}(0)-\sqrt{\kappa}\int_{0}^{t}\exp[-i(\sigma\chi-i\kappa/2)(t-s)]\hat{a}_{{\rm in}}(s)ds,
\end{equation}
where the initial measurement time has been assumed to be zero for simplicity. 

The dispersive coupling of the cavity mode and the qubit causes the qubit state information to be encoded in the output quadrature, 
\begin{equation}
\hat{\mathcal{Z}}_{\rm out}\left(t\right)=\hat{a}_{\rm out}\left(t\right)\exp\left(-i\phi_{h}\right)+\hat{a}^{\dagger}_{\rm out}\left(t\right)\exp\left(i\phi_{h}\right),
\end{equation}
which can be recorded by homodyne detection. Here, $\phi_{h}$ is the measurement angle and $\hat{a}_{\rm out}\left(t\right)=\hat{a}_{\rm in}\left(t\right)+\sqrt{\kappa}\hat{a}\left(t\right)$ refers to the output field of the cavity. The essential parameter quantifying homodyne detection is the SNR. To evaluate the SNR, we typically use the measurement operator,
\begin{equation}\label{eq-measurement-signal}
\hat{M}=\sqrt{\kappa}\int_{0}^{\tau}dt\,\hat{\mathcal{Z}}_{\rm out}\left(t\right),
\end{equation}
with $\tau$ being the measurement time. Its average $\average{\hat{M}}$ corresponds to the qubit-state-dependent signal. The fluctuation noise of the measurement operator $\hat{M}$ is described by
$\hat{M}_{N}=\hat{M}-\average{\hat{M}}$. With these quantities, the SNR is defined as
\begin{equation}\label{seq_SNR_definition}
{\rm SNR}=\frac{\left|\average{\hat{M}} _{\uparrow}-\average{\hat{M}}_{\downarrow}\right|}{\sqrt{\average{\hat{M}_{N}^{2}}_{\uparrow}+\average{\hat{M}_{N}^{2}}_{\downarrow}}},
\end{equation}
where the arrows $\uparrow$ (i.e., $\sigma=+1$) and $\downarrow$ (i.e., $\sigma=-1$) refer to the excited and ground states of the qubit, respectively.

Consider a coherent measurement tone $\average{\hat{a}_{\rm in}\left(t\right)}=\alpha_{\rm in}e^{i\phi_{\rm in}}$. The averaged cavity field can be expressed as
\begin{equation}
\average{a(t)}=i\frac{\sqrt{\kappa}\alpha_{{\rm in}}e^{i\phi_{{\rm in}}}}{\text{\ensuremath{\sigma\chi}}-i\kappa/2}\left\{1-\exp[-i(\sigma\chi-i\kappa/2)t]\right\},
\end{equation}
under the initial condition $\average{\hat{a}(0)}=0$, and the number of cavity photons is accordingly given by
\begin{equation}
	n\left(t\right)=\average{\hat{a}^{\dagger}\left(t\right)\hat{a}\left(t\right)}=\sinh^{2}(r)+\frac{4\alpha_{{\rm in}}^{2}}{\kappa}\cos^{2}(\psi)\left[1+e^{-\kappa t}-2\cos(\chi t)e^{-\kappa t/2}\right],
\end{equation}
where $\tan(\psi)=2\chi/\kappa$. Here, we have assumed that at the initial measurement time $t=0$, the cavity subject to IES is already in a steady state. Then, we find 
\begin{align}\label{seq-signal-separation-injected squeezing}
\average{M} _{\uparrow}-\average{M}_{\downarrow}=\frac{4\alpha_{{\rm in}}}{\sqrt{\kappa}}\sin(2\psi)\sin(\phi_{h}-\phi_{{\rm in}})\left\{ \kappa\tau-4\cos^{2}(\psi)\left[1-\frac{\sin(2\psi+\chi\tau)}{\sin(2\psi)}e^{-\kappa t/2}\right]\right\}.
\end{align}
Note that this expression of the signal separation is the same as that in the standard dispersive readout of no squeezing.

We now derive the measurement noise. The quantum fluctuation operator, $\hat{\mathcal{A}}_{\rm out}(t)=\hat{a}_{\rm out}(t)-\average{\hat{a}_{\rm out}(t)}$, of the output field has the form
\begin{equation}\label{seq-output-fluctuation}
\hat{\mathcal{A}}_{\rm out}(t)=\hat{\mathcal{A}}_{\rm in}(t)+\sqrt{\kappa}\hat{\mathcal{A}}(t).
\end{equation}
Here, $\hat{\mathcal{A}}(t)=\hat{a}(t)-\average{\hat{a}(t)}$ represents the quantum fluctuation of the cavity field, and from Eq.~(\ref{seq-cavity-field-operator-injecting-squeezing}), it is found to be
\begin{equation}\label{seq-cavity-field-fluctuation-injecting-squeezing}
\hat{\mathcal{A}}(t)=\exp[-i(\sigma\chi-i\kappa/2)t]\hat{\mathcal{A}}(0)-\sqrt{\kappa}\int_{0}^{t}\exp[-i(\sigma\chi-i\kappa/2)(t-s)]\hat{\mathcal{A}}_{{\rm in}}(s)ds.
\end{equation}
Since, as assumed above, the cavity subject to IES is already in a steady state at $t=0$, we therefore have:
\begin{align}
\average{\hat{\mathcal{A}}^{\dagger}(0)\hat{\mathcal{A}}(0)}=\;&\sinh^2(r),\quad \quad \quad  \average{\hat{\mathcal{A}}(0)\hat{\mathcal{A}}^{\dagger}(0)}=1+\average{\hat{\mathcal{A}}^{\dagger}(0)\hat{\mathcal{A}}(0)},\\
\average{\hat{\mathcal{A}}(0)\hat{\mathcal{A}}(0)}=\;&\frac{1}{2}e^{i\varphi}\sinh(2r), \quad \average{\hat{\mathcal{A}}^\dagger(0)\hat{\mathcal{A}}^\dagger(0)}=\average{\hat{\mathcal{A}}(0)\hat{\mathcal{A}}(0)}^{*}.
\end{align}
With these initial conditions, we find that the measurement noise, expressed as
\begin{align}\label{eq-meaurement-noise}
\average{\hat{M}_{N}^{2}}=\;&\kappa\int_{0}^{\tau}\int_{0}^{\tau}dt_{1}dt_{2}\bigg\{\average{\hat{\mathcal{A}}_{{\rm out}}(t_{1})\hat{\mathcal{A}}_{{\rm out}}(t_{2})} e^{-i2\phi_{h}}+\average{\hat{\mathcal{A}}_{{\rm out}}^{\dagger}(t_{1})\hat{\mathcal{A}}_{{\rm out}}(t_{2})} \nonumber\\
\;&+\average{\hat{\mathcal{A}}_{{\rm out}}(t_{1})\hat{\mathcal{A}}_{{\rm out}}^{\dagger}(t_{2})}+\average{\hat{\mathcal{A}}_{{\rm out}}^{\dagger}(t_{1})\hat{\mathcal{A}}_{{\rm out}}^{\dagger}(t_{2})} e^{i2\phi_{h}}\bigg\},
\end{align}
is given by
\begin{align}
\average{\hat{M}_{N}^{2}}=\;&\kappa\tau\cosh(2r)+\frac{1}{2}\bigg[3\cos(\varphi-2\phi_{h})-(3-2\kappa\tau)\cos(4\sigma\psi-\varphi+2\phi_{h})\nonumber\\
\;&+6\sin(2\sigma\psi)\sin(4\sigma\psi-\varphi+2\phi_{h})-16e^{-\kappa\tau/2}\cos(\sigma\psi)\sin(2\sigma\psi)\sin(3\sigma\psi-\varphi+2\phi_{h}+\chi\tau)\nonumber\\
\;&+4e^{-\kappa\tau}\cos(\sigma\psi)\sin(2\sigma\psi)\sin(3\sigma\psi-\varphi+2\phi_{h}+2\chi\tau)\bigg]\sinh(2r),
\end{align}
and therefore we obtain
\begin{equation}\label{seq-total-noise-ES}
\average{M_{N}^{2}}_{\downarrow}+\average{M_{N}^{2}}_{\uparrow}	=2\kappa\tau\left[\cosh(2r)+\cos(\varphi-2\phi_{h})\sinh(2r)\mathcal{F}(\tau)\right].
\end{equation}
Here, 
\begin{align}
\mathcal{F}(\tau)=\;&	\frac{1}{2\kappa\tau}\bigg\{3+3\cos(2\psi)-(3-2\kappa\tau)\cos(4\psi)-3\cos(6\psi)\nonumber\\
\;&+4\cos(\psi)\sin(2\psi)\left[e^{-\kappa\tau}\sin(3\psi+2\chi\tau)-4e^{-\kappa\tau/2}\sin(3\psi+\chi\tau)\right]\bigg\}.
\end{align}
It is seen that for a given measurement time $\kappa\tau$, the noise, $\average{M_{N}^{2}}_{\downarrow}+\average{M_{N}^{2}}_{\uparrow}$, can be optimized for $\varphi-2\phi_{h}=\pi$ if $\mathcal{F}(\tau)>0$, or $\varphi-2\phi_{h}=0$ if $\mathcal{F}(\tau)<0$. 

\begin{figure*}[b]
	\centering
	\includegraphics[width=17cm]{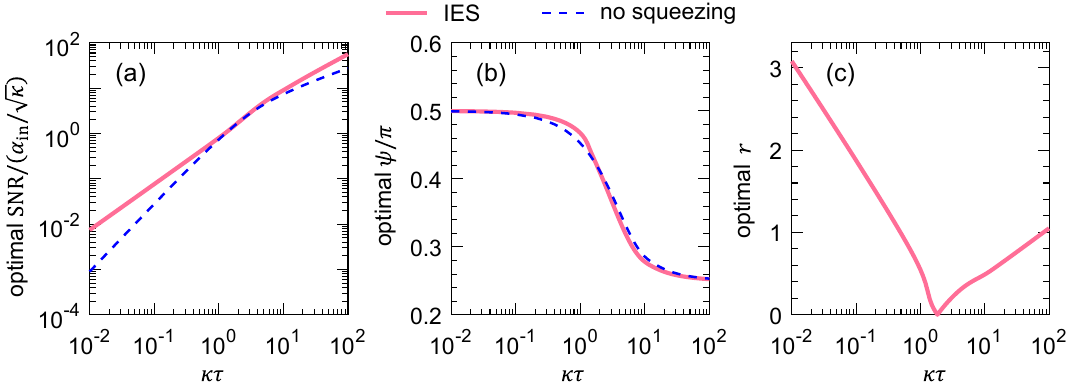}
	\caption{Comparison of DQR with IES (solid curve) and no squeezing (dashed curve). (a) Optimal SNR as a function of the measurement time $\kappa\tau$. (b), (c) Optimal angle $\psi$ and squeezing parameter $r$, corresponding to the optimal SNR in (a).}\label{fig_SNR_ES}
\end{figure*}

\begin{figure*}[t]
	\centering
	\includegraphics[width=16cm]{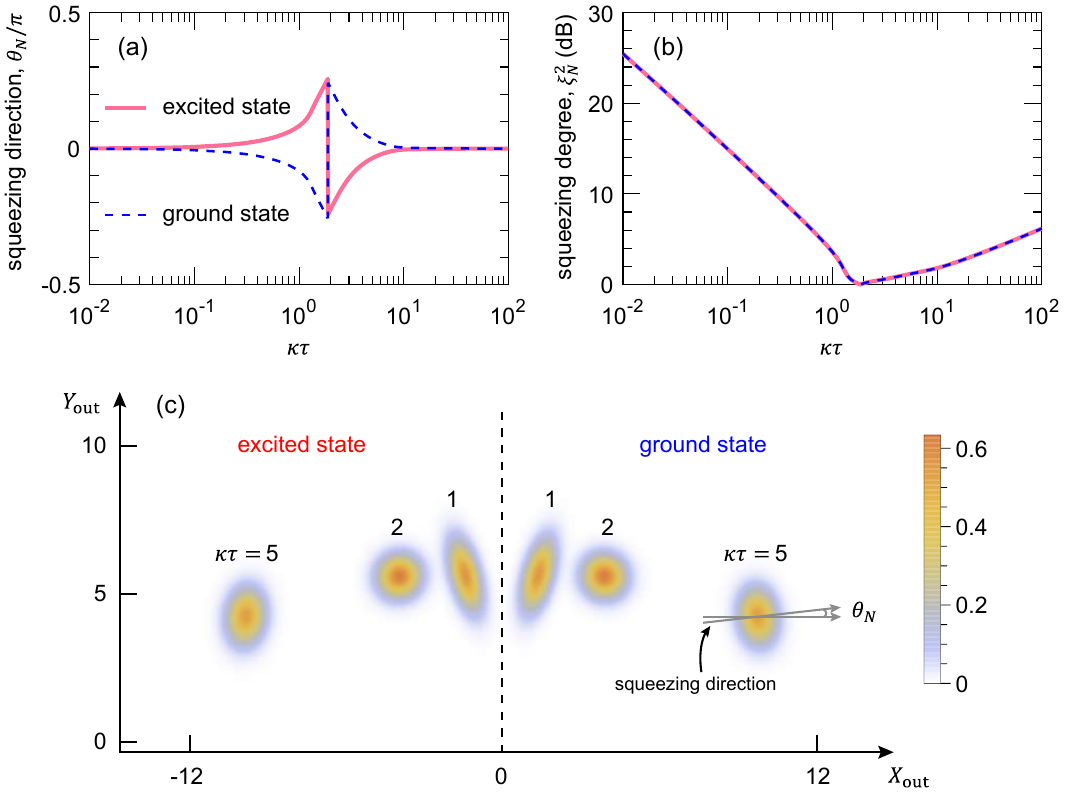}
	\caption{(a) Squeezing direction, (b) squeezing degree, and (c) phase-space representation of the measurement noise $\average{\hat{M}_{N}^{2}}$, corresponding to the optimal SNR in Fig.~\ref{fig_SNR_ES}. Solid (dashed) curves in (a) and (b), and the Wigner functions on the left- (right-) hand side of the vertical dashed line in (c) correspond to the excited (ground) state of the qubit. In (c), we chose three different measurement times, i.e., $\kappa\tau=1$, $2$, $5$, as an example;  $\theta_{N}\in[-\pi/2,\pi/2]$ refers to the angle between the squeezing direction and the horizontal axis (i.e., the measurement direction).}\label{fig_Wigner_function_ES}
\end{figure*}

In Fig.~\ref{fig_SNR_ES}(a), we compare the optimal SNR of DQR using IES to that of the standard case of no squeezing; and the corresponding optimal angle $\psi$ and squeezing parameter $r$ are plotted in Figs.~\ref{fig_SNR_ES}(b) and~\ref{fig_SNR_ES}(c), respectively. It is seen from Fig.~\ref{fig_SNR_ES}(a) that there is almost no improvement in the SNR at a measurement time $\tau\sim1/\kappa$, which is the most interesting experimentally. Note that in the limits of $\kappa\tau\rightarrow0$ and $\kappa\tau\rightarrow\infty$, we can have
\begin{equation}
\average{M_{N}^{2}}_{\downarrow}+\average{M_{N}^{2}}_{\uparrow}	\simeq 2\kappa\tau \exp(-2r),
\end{equation}
which indicates an exponential decrease in the measurement noise, and in turn an exponential increase in the SNR. However, both of these limits are impractical in experiments. In the limit $\kappa\tau\rightarrow0$, the resulting SNR is extremely small, although exponentially increased. As can be seen in Fig.~\ref{fig_SNR_ES}(a), in order to have a significant increase of the SNR, the measurement time $\tau$ needs to be $\sim10^{-2}/\kappa$, at which the amplitude of the measurement tone, $\alpha_{\rm in}$, needs to be $\sim10^{2}\sqrt{\kappa}$ to make the SNR larger than $1$. Such a measurement tone is too strong, and is often not feasible in practice; because it can easily break down the dispersive approximation and, thus, destroy the measurement system. At the same time, as shown in Fig.~\ref{fig_SNR_ES}(b), the qubit-cavity dispersive coupling $\chi$ reaches $\sim10^{2}\kappa$, which is also rather unfortunate due to the fact that how to achieve such a strong nonlinearity in experiments is still an extremely challenging task. In the opposite limit $\kappa\tau\rightarrow\infty$, a significant increase in the SNR needs a measurement time much larger than $10^2/\kappa$, which, clearly, is not desired in experiments.
Hence, using IES alone cannot improve the SNR in a practical manner.

In order to further understand why IES is practically not useful for DQR. In Fig.~\ref{fig_Wigner_function_ES}, we plot the squeezing direction, the squeezing degree, and the phase-space representation of the measurement noise $\average{\hat{M}_{N}^{2}}$ for the ground and excited states of the qubit for the optimal case of Fig.~\ref{fig_SNR_ES}. Here, the squeezing direction is described by an angle $\theta_{N}$ from the horizontal axis, i.e., the measurement direction [see Fig.~\ref{fig_Wigner_function_ES}(c)], and the squeezing degree is defined as 
\begin{equation}
\xi^2_{N}=\frac{\average{\hat{M}_{N}^{2}}}{\kappa\tau}.
\end{equation}
Moreover, following Refs.~\cite{strandberg2021wigner,lu2021propagating}, the Wigner function in phase space is defined as:
\begin{equation}\label{eq_output_wigner_function}
	W\left(X_{\rm out},Y_{\rm out}\right)=\frac{1}{2\pi\sqrt{{\rm Det}\left({\bf D}\right)}}\exp\left(-\frac{1}{2}{\bf G}^{\rm T}{\bf D}^{-1}{\bf G}\right),
\end{equation}
where
\begin{align}
	{\bf G}=\;\left(X_{\rm out}-\average{\hat{X}_{\rm out}},Y_{\rm out}-\average{\hat{Y}_{\rm out}}\right)^{\rm T},
\end{align}
\vspace{-4mm}
\begin{align}
	{\bf D}=\;&\left(
	\begin{array}{cc}
		\average{\hat{X}_{\rm out}^{2}}-\average{\hat{X}_{\rm out}}^{2} & \average{\hat{X}_{\rm out}\hat{Y}_{\rm out}+\hat{Y}_{\rm out}\hat{X}_{\rm out}}/2-\average{\hat{X}_{\rm out}}\average{\hat{Y}_{\rm out}} \\
		\average{\hat{X}_{\rm out}\hat{Y}_{\rm out}+\hat{Y}_{\rm out}\hat{X}_{\rm out}}/2-\average{\hat{X}_{\rm out}}\average{\hat{Y}_{\rm out}}  &  \average{\hat{Y}_{\rm out}^{2}}-\average{\hat{Y}_{\rm out}}^{2} \\
	\end{array}
	\right).
\end{align}
Here,
\begin{align}
	\hat{X}_{\rm out}=\;\frac{1}{2}\left(\hat{A}+\hat{A}^{\dagger}\right),\quad
	\hat{Y}_{\rm out}=\;\frac{1}{2i}\left(\hat{A}-\hat{A}^{\dagger}\right), \quad {\rm and} \quad \hat{A}=\frac{1}{\sqrt{\tau}}\int_{0}^{\tau}dt\;\hat{a}_{s,{\rm out}}\left(t\right).
\end{align}
It can be readily verified that $\left[\hat{A},\hat{A}^{\dagger}\right]=1$ and $\left[\hat{X}_{\rm out},\hat{Y}_{\rm out}\right]=i$.

Clearly, there is a direct correspondence between the results of Fig.~\ref{fig_SNR_ES} and Fig.~\ref{fig_Wigner_function_ES}. It is seen from Fig.~\ref{fig_Wigner_function_ES}(a) that with increasing the measurement time, the squeezing directions of the measurement noises $\average{\hat{M}_{N}^{2}}_{\downarrow}$ and $\average{\hat{M}_{N}^{2}}_{\uparrow}$ are rotated in opposite directions. In the two opposite limits $\kappa\tau\rightarrow0$ and $\kappa\tau\rightarrow\infty$, these two squeezing directions are almost the same, thus giving an exponential but impractical increase in the SNR. However, in the experimentally most interesting regime where $\tau\sim1/\kappa$, there is a large angle between them, as can be seen more clearly in Fig.~\ref{fig_Wigner_function_ES}(c). The presence of such an angle increases the overlap between the two pointer states. In order to reduce this overlap and achieve an optimal SNR, the squeezing degrees of the measurement noises $\average{\hat{M}_{N}^{2}}_{\downarrow}$ and $\average{\hat{M}_{N}^{2}}_{\uparrow}$ have to decrease (even to zero, corresponding to the perpendicular squeezing directions), as plotted in Fig.~\ref{fig_Wigner_function_ES}(b). These competing processes lead to almost no improvement of the SNR. 

\section{Dispersive qubit readout with intracavity squeezing}
\label{Dispersive qubit readout only with intracavity squeezing}
Having discussed the case using IES, we consider in this section DQR with intracavity squeezing (ICS) generated by a two-photon driving. We demonstrate that in the case of using ICS, there also exists a rotation of squeezing similar to the case of using IES; and even worse, the degree of squeezing needs to increase gradually from the zero initial value by increasing the measurement time $\kappa\tau$. {\it Consequently, ICS leads to almost no improvement in the SNR at any measurement time.}

The Hamiltonian for DQR with a two-photon driven cavity reads
\begin{equation}
\hat{H}=\Omega\left[\hat{a}^{2}\exp\left(-i\theta\right)+{\rm H.c.}\right]+\chi \hat{a}^{\dagger}\hat{a}\hat{\sigma}_{z},
\end{equation}
where $\Omega$ and $\theta$ are the amplitude and phase of the two-photon driving, respectively.
The Langevin equation of motion for the cavity mode $\hat{a}$ is accordingly given by
\begin{equation}\label{seq_equation_motion_a}
\dot{\hat{a}}(t)=-i\left(\sigma\chi-i\kappa/2\right)\hat{a}-i2\Omega\exp\left(i\theta\right) \hat{a}^{\dagger}-\sqrt{\kappa}\hat{a}_{\rm in}\left(t\right).
\end{equation}
Moreover, the correlations for the noise operator $\hat{\mathcal{A}}_{\rm in}\left(t\right)=\hat{a}_{\rm in}\left(t\right)-\average{\hat{a}_{\rm in}\left(t\right)}$ are:
\begin{align}
\average{\hat{\mathcal{A}}_{\rm in}\left(t\right)\hat{\mathcal{A}}^{\dagger}_{\rm in}\left(t^{\prime}\right)}=\;&[\hat{\mathcal{A}}_{\rm in}\left(t\right),\hat{\mathcal{A}}^{\dagger}_{\rm in}\left(t^{\prime}\right)]=\delta\left(t-t^{\prime}\right),\\
\average{\hat{\mathcal{A}}_{\rm in}^{\dagger}\left(t\right)\hat{\mathcal{A}}_{\rm in}\left(t^{\prime}\right)}=\;&\average{\hat{\mathcal{A}}_{\rm in}\left(t\right)\hat{\mathcal{A}}_{\rm in}\left(t^{\prime}\right)}=0.
\end{align}
According to Eq.~(\ref{seq_equation_motion_a}), the cavity mode $\hat{a}$ is found to be
\begin{align}\label{seq_evolution_a}
\hat{a}\left(t\right)=\;&\Lambda(t)a(0)-\Gamma(t)a^{\dagger}(0)\nonumber\\
\;&-\sqrt{\kappa}\int_{0}^{t}ds\Lambda(t-s)a_{{\rm in}}(s)+\sqrt{\kappa}\int_{0}^{t}ds\Gamma(t-s)a_{{\rm in}}^{\dagger}(s),
\end{align}
where 
\begin{align}
\Lambda(t)=\;&\frac{1}{\lambda}\left[\lambda\cos(\lambda t)-i\sigma\chi\sin(\lambda t)\right]\exp(-\kappa t/2),\\
\Gamma(t)=\;&\frac{2}{\lambda}ie^{i\theta}\Omega\sin(\lambda t)\exp(-\kappa t/2),\\
\lambda=\;&\sqrt{\chi^{2}-4\Omega^{2}}.
\end{align}
As a direct result of Eq.~(\ref{seq_evolution_a}), the averaged cavity field is given, under the initial condition of $\average{\hat{a}\left(0\right)}=0$, by
\begin{align}\label{seq_average_intracavity_field_full}
\average{\hat{a}\left(t\right)}=\;&\frac{2\sqrt{\kappa}\alpha_{{\rm in}}}{\kappa^{2}+4\lambda^{2}}\bigg\{ i4\Omega e^{i(\theta-\phi_{{\rm in}})}-(\kappa-i2\sigma\chi)e^{i\phi_{{\rm in}}}\nonumber\\
\;&-\frac{1}{\lambda}\left[(2\lambda^{2}+i\kappa\sigma\chi)e^{i\phi_{{\rm in}}}+i2\Omega\kappa e^{i(\theta-\phi_{{\rm in}})}\right]\sin(\lambda t)e^{-\kappa t/2}\nonumber\\
\;&+\left[(\kappa-i2\sigma\chi)e^{i\phi_{{\rm in}}}-i4\Omega e^{i(\theta-\phi_{{\rm in}})}\right]\cos(\lambda t)e^{-\kappa t/2}\bigg\}.
\end{align}
It is seen that in order to stabilize the
system, we need to restrict our discussions to the case either when $\lambda$ is a real number (i.e., $\chi>2\Omega$) or an imaginary number but with $|\lambda|<\kappa/2$. It then follows, according to the input-output relation $\hat{a}_{\rm out}\left(t\right)=\hat{a}_{\rm in}\left(t\right)+\sqrt{\kappa}\hat{a}\left(t\right)$, that 
\begin{equation}\label{seq-signal-separation-intracavity squeezing}
\average{\hat{M}} _{\uparrow}-\average{\hat{M}}_{\downarrow}=\frac{16(\chi/\kappa)\alpha_{{\rm in}}}{\sqrt{\kappa}}\cos^{2}(\psi)\sin(\phi_{h}-\phi_{{\rm in}})\left\{ \kappa\tau-4\cos^{2}(\psi)\left[1-\frac{\sin(2\psi+\lambda\tau)}{\sin(2\psi)}e^{-\kappa\tau/2}\right]\right\}.
\end{equation}
Note that here, we have defined $\tan(\psi)=2\lambda/\kappa$, instead of $\tan(\psi)=2\chi/\kappa$ as in Sec.~\ref{Dispersive qubit readout only with injected squeezing}. 

\begin{figure*}[b]
	\centering
	\includegraphics[width=17cm]{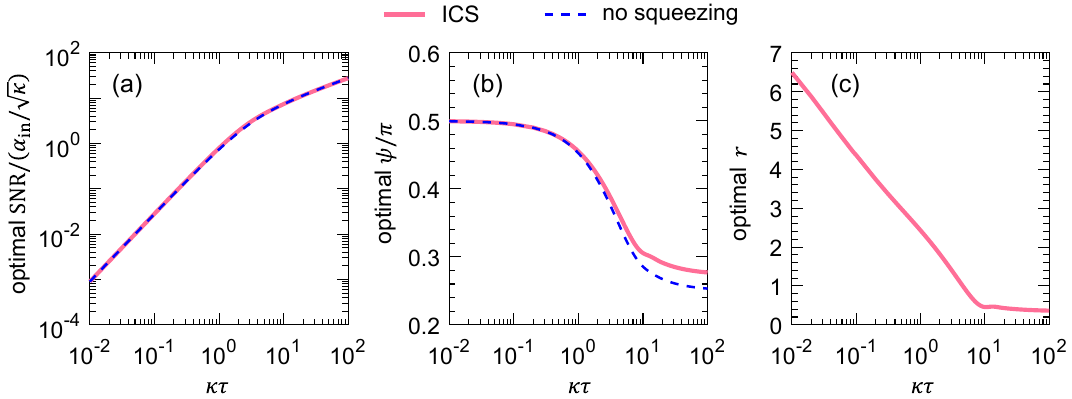}
	\caption{Comparison of DQR with ICS (solid curve) and no squeezing (dashed curve). (a) Optimal SNR as a function of the measurement time $\kappa\tau$. (b), (c) Optimal angle $\psi$ and squeezing parameter $r$, corresponding to the optimal SNR in (a). In (b), $\tan(\psi)=2\lambda/\kappa$ for the readout with ICS, but $\tan(\psi)=2\chi/\kappa$ for the standard case of no squeezing.}\label{fig_SNR_IS}
\end{figure*}
\begin{figure*}[t]
	\centering
	\includegraphics[width=16cm]{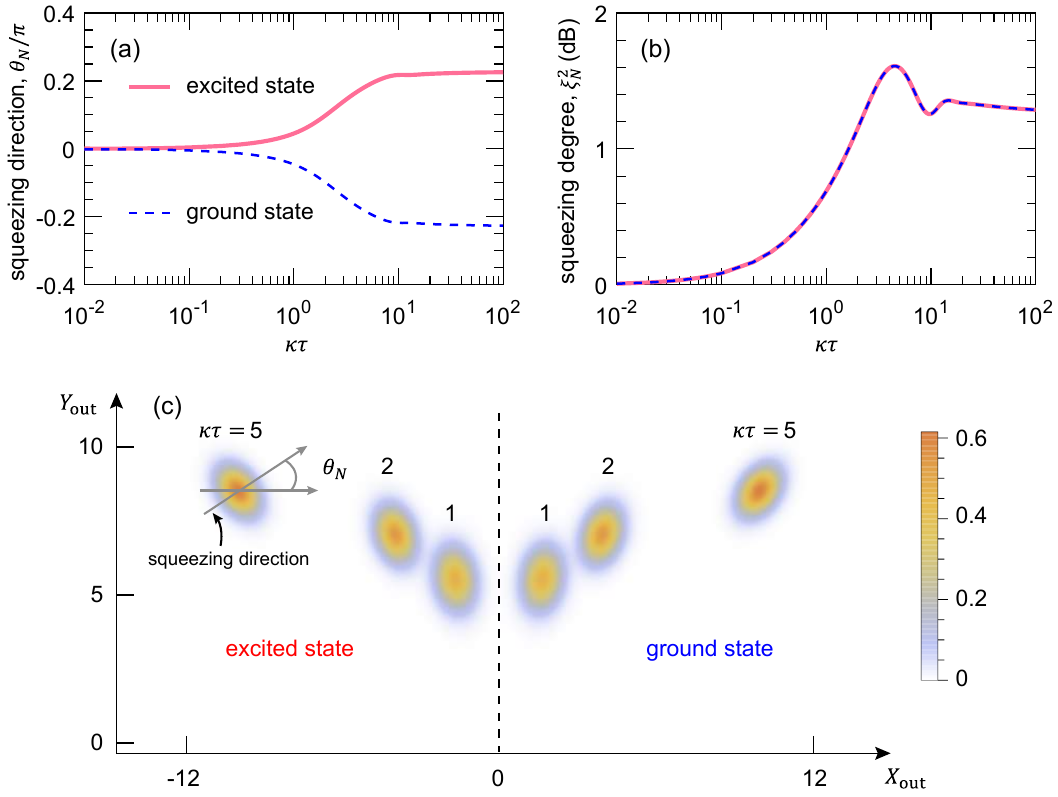}
	\caption{(a) Squeezing direction, (b) squeezing degree, and (c) phase-space representation of the measurement noise $\average{\hat{M}_{N}^{2}}$, corresponding to the optimal SNR in Fig.~\ref{fig_SNR_IS}. Solid (dashed) curves in (a) and (b), and the Wigner functions on the left- (right-) hand side of the vertical dashed line in (c) correspond to the excited (ground) state of the qubit. In (c), we chose three different measurement times, i.e., $\kappa\tau=1$, $2$, $5$, as an example;  $\theta_{N}\in[-\pi/2,\pi/2]$ refers to the angle between the squeezing direction and the horizontal axis (i.e., the measurement direction).}\label{fig_Wigner_function_IS}
\end{figure*}
Furthermore, the quantum fluctuation operator of the cavity field $\hat{a}(t)$ is given by
\begin{equation}
\hat{\mathcal{A}}(t)=\Lambda(t)\hat{\mathcal{A}}(0)-\Gamma(t)\hat{\mathcal{A}}^{\dagger}(0)-\sqrt{\kappa}\int_{0}^{t}ds\Lambda(t-s)\hat{\mathcal{A}}_{\rm in}(s)+\sqrt{\kappa}\int_{0}^{t}ds\Gamma(t-s)\hat{\mathcal{A}}_{\rm in}^{\dagger}(s),
\end{equation}
according to Eq.~(\ref{seq_evolution_a}). We further assume that the two-photon driven cavity is already in a steady state at the initial measurement time $t=0$. Under this assumption, the correlations for the cavity-field noise operator $\hat{\mathcal{A}}(0)$ read:
\begin{align}
\average{\hat{\mathcal{A}}^{\dagger}(0)\hat{\mathcal{A}}(0)}=\;&\frac{8\Omega^{2}}{\kappa^{2}-16\Omega^{2}},\quad \quad \quad \quad \average{\hat{\mathcal{A}}(0)\hat{\mathcal{A}}^{\dagger}(0)}=1+\average{\hat{\mathcal{A}}^{\dagger}(0)\hat{\mathcal{A}}(0)},\\
\average{\hat{\mathcal{A}}(0)\hat{\mathcal{A}}(0)}=\;&-ie^{i\theta}\frac{2\kappa\Omega}{\kappa^{2}-16\Omega^{2}}, \quad \average{\hat{\mathcal{A}}^\dagger(0)\hat{\mathcal{A}}^\dagger(0)}=\average{\hat{\mathcal{A}}(0)\hat{\mathcal{A}}(0)}^{*}.
\end{align}
Then after a straightforward but tedious calculation, we find the measurement noise to be
\begin{equation}
\average{\hat{M}_{N}^{2}}=\mathcal{G}_{0}(\tau)-\sin(2\phi_{h}-\theta)\mathcal{G}_{s}(\tau)+\frac{\sigma\chi}{\kappa}\cos(2\phi_{h}-\theta)\mathcal{G}_{c}(\tau),
\end{equation}
where
\begin{align}
\mathcal{G}_{0}(\tau)=\;&\frac{1}{2}\kappa\tau\left\{ 1+\cosh(r)+\left[5+8\cos(2\psi)+2\cos(4\psi)-\cosh(r)\right]\tanh^{2}(\frac{r}{2})\right\}\nonumber\\
\;&-2\cos^{2}(\psi)\left\{5+3\cos(4\psi)+\cos(2\psi)\left[9-2\cosh(r)\right]-3\cosh(r)\right\} \tanh^{2}(\frac{r}{2})\nonumber\\
\;&-e^{-\kappa\tau}\left[2-\cos(2\psi+2\lambda\tau)-\cos(4\psi+2\lambda\tau)\right]\left[\cos(2\psi)-\cosh(r)\right]\cot^{2}(\psi)\tanh^{2}(\frac{r}{2})\nonumber\\
\;&-8e^{-\kappa\tau/2}\cos^{2}(\psi)\tanh^{2}(\frac{r}{2})\bigg\{ \left[\cos(\lambda\tau)-\cot(\psi)\sin(4\psi+\lambda\tau)\right]\cosh^{2}(\frac{r}{2})\nonumber\\
\;&+4\cos^{2}(\psi)\cot(\psi)\sin(2\psi+\lambda\tau)\sinh^{2}(\frac{r}{2})\bigg\},
\end{align}
\begin{align}
\mathcal{G}_{s}(\tau)=\;&2\cos^{2}(\psi)\left\{ -1-3\cos(4\psi)+\cosh(r)+\cos(2\psi)\left[-3+2\kappa\tau+2\cosh(r)\right]\right\} \tanh(\frac{r}{2})\nonumber\\
\;&-2e^{-\kappa\tau}\cos(\psi)\cot(\psi)\sin(3\psi+2\lambda\tau)\left[\cos(2\psi)-\cosh(r)\right]\tanh(\frac{r}{2})\nonumber\\
\;&-4e^{-\kappa\tau/2}\cos(\psi)\cot(\psi)\left[\sin(3\psi+\lambda\tau)\sinh(r)-2\cos(\psi)\sin(4\psi+\lambda\tau)\tanh(\frac{r}{2})\right],
\end{align}
\begin{align}
\mathcal{G}_{c}(\tau)=\;&8\cos^{4}(\psi)\left[3-2\kappa\tau+6\cos(2\psi)-2\cosh(r)\right]\tanh(\frac{r}{2})\nonumber\\
\;&-16e^{-\kappa\tau/2}\cos^{4}(\psi)\cot(\psi)\sinh^{2}(\frac{r}{2})\left[\coth(\frac{r}{2})\sec^{2}(\psi)\sin(4\psi+\lambda\tau)-4\sin(2\psi+\lambda\tau)\tanh(\frac{r}{2})\right]\nonumber\\
\;&+8e^{-\kappa\tau}\cos^{2}(\psi)\sinh(\frac{r}{2})\bigg\{ \cos(\psi)\cos(3\psi+2\lambda\tau)\cosh(\frac{r}{2})\nonumber\\
\;&-\left[1-\cos(\psi)\cos(3\psi+2\lambda\tau)\right]\cot^{2}(\psi)\sinh(\frac{r}{2})\tanh(\frac{r}{2})\bigg\}.
\end{align}
Here, we have defined a squeezing parameter,
\begin{equation}
r=\ln\left(\frac{\kappa+4\Omega}{\kappa-4\Omega}\right),
\end{equation}
which, in fact, determines the squeezing degree of the output field of the two-photon driven cavity in the absence of the qubit. Consequently, we have
\begin{equation}
\average{\hat{M}_{N}^{2}}_{\downarrow}+\average{\hat{M}_{N}^{2}}_{\uparrow}=2\mathcal{G}_{0}(\tau)-2\sin(2\phi_{h}-\theta)\mathcal{G}_{s}(\tau).
\end{equation}
It is seen that for a given measurement time $\kappa\tau$, the noise, $\average{\hat{M}_{N}^{2}}_{\downarrow}+\average{\hat{M}_{N}^{2}}_{\uparrow}$, can be optimized for $2\phi_{h}-\theta=\pi/2$ if $\mathcal{G}_{s}(\tau)>0$, or for $2\phi_{h}-\theta=-\pi/2$ if $\mathcal{G}_{s}(\tau)<0$. The number of cavity photons is accordingly given by
\begin{equation}
	n\left(t\right)=\average{\hat{a}^{\dagger}\left(t\right)\hat{a}\left(t\right)}\nonumber\\
	=\frac{1}{8}\left[ 4\cos^{2}(\psi)-e^{-\kappa t}\mathcal{Q}_{0}\right]\tanh^{2}(\frac{r}{2})+\left(\frac{\alpha_{{\rm in}}}{\sqrt{\kappa}}\right)^{2}\mathcal{Q}_{1},
\end{equation}
where
\begin{equation}
	\mathcal{Q}_{0}=\left[2-\cos(2\lambda t)-\cos(2\psi+2\lambda t)\right]\left[\cos(2\psi)-\cosh(r)\right]\csc^{2}(\psi),
\end{equation}
\begin{align}
	\mathcal{Q}_{1}=&4\left(\frac{\alpha_{{\rm in}}}{\sqrt{\kappa}}\right)^{2}\cos^{2}(\psi)\bigg\{ 1+e^{-\kappa t}-2e^{-\kappa t/2}\cos(\lambda t)\nonumber\\
	&+\left[\sin(2\psi)-2e^{-\kappa t/2}\sin(2\psi+\lambda t)+e^{-\kappa t}\sin(2\psi+2\lambda t)\right]\cot(\psi)\tanh(\frac{r}{2})\nonumber\\
	&+2\left[\cos(\psi)-e^{-\kappa t/2}\cot(\psi)\sin(\psi+\lambda t)\right]^{2}\tanh^{2}(\frac{r}{2})\bigg\}.	
\end{align}

In Fig.~\ref{fig_SNR_IS}(a), we compare the optimal SNR of DQR using ICS (i.e., using a two-photon driven cavity) to that of the standard case of no squeezing; and the corresponding optimal angle $\psi$ and squeezing parameter $r$ are plotted in Figs.~\ref{fig_SNR_IS}(b) and~\ref{fig_SNR_IS}(c), respectively. It is seen that there is almost no improvement in the SNR for any measurement time. 

We now discuss the physical reasons why the SNR can hardly be improved by ICS. In analogy to the analysis of the case of using IES in Sec.~\ref{Dispersive qubit readout only with injected squeezing}, we plot in Fig.~\ref{fig_Wigner_function_IS} the squeezing direction $\theta_{N}$, the squeezing degree $\xi^2_{N}$, and the phase-space representation of the measurement noise $\average{\hat{M}_{N}^{2}}$ for the ground and excited states of the qubit for the optimal case of Fig.~\ref{fig_SNR_IS}. We find from Figs.~\ref{fig_Wigner_function_IS}(a) and~\ref{fig_Wigner_function_IS}(b) that, when $\kappa\tau\rightarrow0$, the squeezing directions of the measurement noises $\average{\hat{M}_{N}^{2}}_{\downarrow}$ and $\average{\hat{M}_{N}^{2}}_{\uparrow}$ are almost the same, but at the same time, their squeezing degrees are extremely weak. Moreover, as $\kappa\tau$ increases, the squeezing degrees are increased and gradually converged to a value of $\simeq1.27$~dB in the limit $\kappa\tau\rightarrow\infty$; but at the same time, the squeezing directions are rotated in opposite directions as can be seen more clearly in Fig.~\ref{fig_Wigner_function_IS}(c), and they even become mutually perpendicular in the limit $\kappa\tau\rightarrow\infty$. These features prevent the SNR improvement from using ICS.  

\section{Qubit-cavity dispersive coupling enhanced by squeezing}
\label{Enhanced dispersive coupling}
The $\hat{\sigma}_z$ term in Eq.~(5) in the main article corresponds to an enhanced dispersive coupling between the qubit and the Bogoliubov mode $\hat{\beta}$, which is of the form 
\begin{equation}\label{eq-dispersive-beta-qubit}
	\hat{V}_{\rm sq}=\chi_{\rm sq}\hat{\beta}^{\dagger}\hat{\beta}\hat{\sigma}_z,
\end{equation}
where $\chi_{\rm sq}$ is the dispersive coupling strength given by
\begin{equation}\label{eq-enhanced-dispersive-coupling}
	\chi_{\rm sq}=g^{2}\left[\frac{\cosh^{2}(r)}{\Delta_{q}-\omega_{{\rm sq}}}+\frac{\sinh^{2}(r)}{\Delta_{q}+\omega_{{\rm sq}}}\right]=\chi\left[\cosh(r)+\frac{\sinh^{2}(r)}{\cosh(r)+2\omega_{\rm sq}\epsilon/g}\right].
\end{equation}
Here, we have defined 
\begin{equation}\label{eq-chi}
	\chi=g\epsilon,
\end{equation}
with 
\begin{equation}\label{eq-epsilon}
	\epsilon=\frac{g\cosh(r)}{\Delta_{q}-\omega_{\rm sq}}.
\end{equation}
Note that a very recent experiment~\cite{villiers2024dynamically} in superconducting circuits has demonstrated this enhanced dispersive coupling and, particularly, the increase of $\chi_{\rm sq}$ with the squeezing parameter $r$.

In order to better understand the physical meaning of the dispersive coupling $\hat{V}_{\rm sq}$, we now present its detailed derivation. We begin with the full Hamiltonian in Eq.~(1) in the main article and, for convenience, we reproduce it here,
\begin{equation}\label{eq-full-Hamiltonian}
	\hat{H}=\Delta_{c}\hat{a}^{\dagger}\hat{a}+\frac{1}{2}\Delta_{q}\hat{\sigma}_{z}+g\left(\hat{a}^{\dagger}\hat{\sigma}_{-}+\hat{a}\hat{\sigma}_{+}\right)+\Omega\left(e^{i\theta}\hat{a}^{\dagger2}+e^{-i\theta}\hat{a}^{2}\right).
\end{equation}
Upon introducing a Bogoliubov mode 
\begin{equation}\label{eq-Bogoliubov-transformation}
	\hat{\beta}=\cosh(r_c)\hat{a}+e^{i\theta}\sinh(r_c)\hat{a}^{\dagger},
\end{equation}
with $\tanh(2r_c)=2\Omega/\Delta_{c}$, the cavity Hamiltonian is diagonalized, yielding
\begin{equation}
	\Delta_{c}\hat{a}^{\dagger}\hat{a}+\Omega\left(e^{i\theta}\hat{a}^{\dagger2}+e^{-i\theta}\hat{a}^{2}\right)=\omega_{\rm sq}\hat{\beta}^{\dagger}\hat{\beta},
\end{equation}
where $\omega_{\rm sq}=\sqrt{\Delta_{c}^{2}-4\Omega^{2}}$ is the resonance frequency of the mode $\hat{\beta}$. Expressed in terms of the mode $\hat{\beta}$, the full Hamiltonian $\hat{H}$ is then transformed to
\begin{equation}\label{eq-full-Hamiltonian-beta}
	\hat{H}=\omega_{\rm sq}\hat{\beta}^{\dagger}\hat{\beta}+\frac{1}{2}\Delta_{q}\hat{\sigma}_{z}+g\cosh(r)\left(\hat{\beta}^{\dagger}\hat{\sigma}_{-}+\hat{\beta}\hat{\sigma}_{+}\right)-g\sinh(r)\left(e^{-i\theta}\hat{\beta}\hat{\sigma}_{-}+e^{i\theta}\hat{\beta}^{\dagger}\hat{\sigma}_{+}\right).
\end{equation}
Here, we have made a replacement $r_{c}\rightarrow r$, since $r$ denotes a squeezing parameter of IES used in our proposal and, in order to improve the SNR, we need to set $r_c=r$ [see Eq.~(\ref{seq-parameter-condition}) below].

Furthermore, we work within the regime 
\begin{equation}\label{eq-dispersive-regime01}
	\epsilon \ll 1, 
\end{equation} 
so that we can make a dispersive approximation~\cite{gamel2010time}, and then obtain the dispersive coupling $\hat{V}_{\rm sq}$ in Eq.~(\ref{eq-dispersive-beta-qubit}). Note that here, to make the dispersive approximation, we also require to satisfy the condition
\begin{equation}\label{eq-condition02}
	\left(\Delta_{q}+\omega_{\rm sq}\right)\gg g\sinh(r),
\end{equation} 
in addition to $\epsilon\ll1$. 
But since the condition in Eq.~(\ref{eq-condition02}) is certainly satisfied once $\epsilon\ll1$ holds, we can therefore only consider the condition of $\epsilon\ll1$ for the dispersive approximation.

Below, we compare the dispersive coupling $\hat{V}_{\rm sq}$ and the original dispersive coupling, with no squeezing, of the qubit and the bare cavity mode $\hat{a}$, and explain the reason why the resulting improvement in the dispersive coupling is real.

Let us first consider the original dispersive coupling in the absence of squeezing. For clarity, we begin with the Jaynes-Cummings Hamiltonian of a qubit coupled to a cavity mode, i.e.,
\begin{equation}\label{eq-JC-model}
	\hat{H}_{0}=\omega_{c}\hat{a}^{\dagger}\hat{a}+\frac{1}{2}\omega_{q}\hat{\sigma}_{z}+g\left(\hat{a}^{\dagger}\hat{\sigma}_{-}+\hat{a}\hat{\sigma}_{+}\right).
\end{equation}
In the regime where
\begin{equation}\label{eq-dispersive-regime02}
	\epsilon_{0}=\frac{g}{\omega_{q}-\omega_{c}}\ll1,
\end{equation}
we can make a dispersive approximation~\cite{gamel2010time}, yielding a dispersive coupling of the qubit and the cavity mode,
\begin{equation}
	\hat{V}_{0}=\chi_{0}\hat{a}^{\dagger}\hat{a}\hat{\sigma}_{z},
\end{equation}
where $\chi_{0}$ is the dispersive coupling strength given by
\begin{equation}\label{eq-chi0}
	\chi_{0}=\frac{g^2}{\omega_{q}-\omega_{c}}=g\epsilon_{0}.
\end{equation} 
The use of the Jaynes-Cummings interaction given in Eq.~(\ref{eq-JC-model}) is the simplest and most common way to achieve the dispersive coupling $\hat{V}_{0}$.

Note that $\epsilon_{0}$ is a key parameter, which determines the validity or accuracy of the dispersive approximation applied to Eq.~(\ref{eq-JC-model}).
Analogously, as described above, the parameter $\epsilon$ in Eq.~(\ref{eq-epsilon}) determines the validity or accuracy of the dispersive approximation applied to Eq.~(\ref{eq-full-Hamiltonian-beta}), and it plays a role similar to the parameter $\epsilon_{0}$.

Thus, in order to ensure a fair comparison between the dispersive couplings $\hat{V}_{0}$ and $\hat{V}_{\rm sq}$, we need to assume
	\begin{equation}
		\epsilon_{0}=\epsilon,
	\end{equation}
	such that the dispersive approximations applied for $\hat{V}_{0}$ and $\hat{V}_{\rm sq}$, respectively, can have the same validity or accuracy. In such a case, by comparing Eqs.~(\ref{eq-chi}) and~(\ref{eq-chi0}), we see that
	\begin{equation}
		\chi_{0}=\chi.
	\end{equation}
	That is, {\it under the condition that the two dispersive couplings $\hat{V}_{0}$ and $\hat{V}_{\rm sq}$ have the same validity or accuracy, the parameter $\chi$ can be regarded as the original dispersive coupling strength $\chi_{0}$}. Consequently, according to Eq.~(\ref{eq-enhanced-dispersive-coupling}), our dispersive coupling strength $\chi_{\rm sq}$ can be regarded as being enhanced by a factor of 
\begin{equation}
	\cosh(r)+\frac{\sinh^{2}(r)}{\cosh(r)+2\omega_{\rm sq}\epsilon/g},
\end{equation}
compared to the original dispersive coupling strength $\chi_{0}$ (i.e., $\chi$). Moreover, as long as $\epsilon\ll1$, we have
\begin{equation}
	\cosh(r)+\frac{\sinh^{2}(r)}{\cosh(r)+2\omega_{\rm sq}\epsilon/g}\simeq \exp(r)
\end{equation}
and as a result, an exponential enhancement,
\begin{equation}\label{eq-exponential-chisq}
	\chi_{\rm sq}\simeq\chi_{0}\exp(r), \quad {\rm i.e.,} \quad \chi_{\rm sq}\simeq\chi\exp(r).
\end{equation}
Hence, according to the above discussions, our proposal leads to a real improvement in the dispersive coupling and as a result also in the SNR.

Note that in order to obtain the enhanced dispersive coupling $\hat{V}_{\rm sq}$ from Eq.~(\ref{eq-full-Hamiltonian-beta}), we assume that the coupling strengths $g\cosh(r)$ and $g\sinh(r)$ are small, compared to the detunings $\Delta_{q}-\omega_{\rm sq}$ and $\Delta_{q}+\omega_{\rm sq}$, respectively; i.e., $\epsilon\ll1$ [note that, as mentioned above, $\epsilon\ll1$ ensures $\left(\Delta_{q}+\omega_{\rm sq}\right)\gg g\sinh(r)$]. However, in analogy, obtaining the original dispersive coupling $\hat{V}_{0}$ from Eq.~(\ref{eq-JC-model}) is also based on assuming the coupling strength $g$ to be small compared to the detuning $\omega_{q}-\omega_{c}$; i.e., $\epsilon_{0}\ll1$. To have a fair comparison, as mentioned above, we need to set $\epsilon_{0}=\epsilon$. Thus, in this case, the assumption of $\epsilon\ll1$ does not affect the improvement of the dispersive coupling.

\section{Improved dispersive qubit readout with both injected external squeezing and intracavity squeezing}
\label{Enhance dispersive qubit readout}
In this section, we consider the case when IES and ICS are used simultaneously for DQR, and demonstrate that {\it for any measurement time, squeezing in this case can enable an exponential increase of the readout SNR. In particular, the short-time SNR can be increased exponentially with twice the squeezing parameter.} This is in stark contrast to the case of using IES or ICS alone. 

To begin, we consider the Langevin equation of motion,
\begin{equation}\label{eq-equation-of-motion-beta}
\dot{\hat{\beta}}(t)=-i(\omega_{\sigma}-i\frac{\kappa}{2})\hat{\beta}-\sqrt{\kappa}\hat{\beta}_{{\rm in}}(t),
\end{equation}
where $\omega_{\sigma}=\omega_{{\rm sq}}+\sigma\chi_{\rm sq}$, and $\hat{\beta}_{{\rm in}}(t)$ denotes the input field of the Bogoliubov mode $\hat{\beta}$. It is seen that the information about the qubit state is mapped onto the mode $\hat{\beta}$, rather than the bare mode $\hat{a}$. The quantum fluctuation operator, $\hat{\mathcal{B}}_{\rm in}(t)=\hat{\beta}_{{\rm in}}(t)-\average{\hat{\beta}_{{\rm in}}(t)}$, of the input field $\hat{\beta}_{{\rm in}}(t)$, is given by
\begin{equation}
\hat{\mathcal{B}}_{\rm in}(t)=\cosh(r_c)\hat{\mathcal{A}}_{{\rm in}}(t)+e^{i\theta}\sinh(r_c)\hat{\mathcal{A}}_{{\rm in}}^{\dagger}(t),
\end{equation}
according to the Bogoliubov transformation,
\begin{equation}\label{eq-bogoliubov-transformation-beta-in}
\hat{\beta}_{\rm in}(t)=\cosh(r_c)\hat{a}_{{\rm in}}(t)+e^{i\theta}\sinh(r_c)\hat{a}_{{\rm in}}^{\dagger}(t).
\end{equation}
The correlations for $\hat{\mathcal{A}}_{{\rm in}}(t)$ are given in Eqs.~(\ref{sq-correlations-injected-squeezing-01}) and~(\ref{sq-correlations-injected-squeezing-02}) and thus, the correlations for the operator $\hat{\mathcal{B}}_{\rm in}(t)$ are found to be:
\begin{align}
\average{\mathcal{B}_{{\rm in}}^{\dagger}(t)\mathcal{B}_{{\rm in}}(t^{\prime})}=\;&\mathcal{N}\delta(t-t^{\prime}),\quad \quad
\average{\mathcal{B}_{{\rm in}}(t)\mathcal{B}_{{\rm in}}^{\dagger}(t^{\prime})}=\;\left(\mathcal{N}+1\right)\delta(t-t^{\prime}),\\ \average{\mathcal{B}_{{\rm in}}(t)\mathcal{B}_{{\rm in}}(t^{\prime})}=\;&\mathcal{M}\delta(t-t^{\prime}),\quad \quad
\average{\mathcal{B}_{{\rm in}}^{\dagger}(t)\mathcal{B}_{{\rm in}}^{\dagger}(t^{\prime})} =\;\mathcal{M}^{*}\delta(t-t^{\prime}),
\end{align}
where 
\begin{align}
\mathcal{N}=\;&\cosh^{2}(r_c)\sinh^{2}(r)+\sinh^{2}(r_c)\cosh^{2}(r)+\frac{1}{2}\cos(\theta-\varphi)\sinh(2r_c)\sinh(2r),\nonumber\\
\mathcal{M}=\;&\frac{1}{2}\bigg[e^{i\varphi}\cosh^{2}(r_c)\sinh(2r)+e^{i\theta}\sinh(2r_c)\sinh^{2}(r)\nonumber\\
\;&+e^{i\theta}\sinh(2r_c)\cosh^{2}(r)+e^{i(2\theta-\varphi)}\sinh^{2}(r_c)\sinh(2r)\bigg].
\end{align}
This indicates that the mode $\hat{\beta}$ suffers from thermal noise, characterized by $\mathcal{N}$, and two-photon correlation noise, characterized by $\mathcal{M}$. These two types of noise are undesired in our proposal, but having
\begin{equation}\label{seq-parameter-condition}
r_c=r, \quad {\rm and} \quad \theta-\varphi=\pi
\end{equation}
can eliminate them completely, i.e.,
\begin{equation}
	\mathcal{N}=\mathcal{M}=0,
\end{equation}
so that the mode $\hat{\beta}$ suffers only from a simple vacuum noise, i.e., 
\begin{gather}
		\average{\mathcal{B}_{{\rm in}}(t)\mathcal{B}_{{\rm in}}^{\dagger}(t^{\prime})}=\;\delta(t-t^{\prime}),\\
	\average{\mathcal{B}_{{\rm in}}^{\dagger}(t)\mathcal{B}_{{\rm in}}(t^{\prime})}=\average{\mathcal{B}_{{\rm in}}(t)\mathcal{B}_{{\rm in}}(t^{\prime})}=\average{\mathcal{B}_{{\rm in}}^{\dagger}(t)\mathcal{B}_{{\rm in}}^{\dagger}(t^{\prime})}=0.
\end{gather}
In this case, we show below that the measurement noise of the readout can be exponentially suppressed. 

As usual, we formally integrate the equation of motion in Eq.~(\ref{eq-equation-of-motion-beta}) to yield
\begin{equation}
\hat{\beta}(t)=\hat{\beta}(0)\exp[-i(\omega_{\sigma}-i\kappa/2)t]-\sqrt{\kappa}\int_{0}^{t}ds\exp[-i(\omega_{\sigma}-i\kappa/2)(t-s)]\hat{\beta}_{{\rm in}}(s),
\end{equation}
and accordingly, the number of cavity photons in the mode $\hat{\beta}$ is found to be
\begin{equation}
n(t)=\average{\hat{\beta}(t)^\dagger\hat{\beta}(t)}=4|\average{\hat{\beta}_{{\rm in}}(t)}|^2\cos^2(\psi_{\sigma})\left[1+e^{-\kappa t}-2e^{-\kappa t/2}\cos(\omega_{\sigma}t)\right],
\end{equation}
where $\tan(\psi_{\sigma})=2\omega_{\sigma}/\kappa$.
Then, according to the input-output relation $\hat{\beta}_{{\rm out}}(t)=\hat{\beta}_{{\rm in}}(t)+\sqrt{\kappa}\hat{\beta}(t)$, we have
\begin{equation}\label{eq-output-field-beta}
\hat{\beta}_{{\rm out}}(t)=\hat{\beta}_{{\rm in}}(t)+\sqrt{\kappa}\hat{\beta}(0)\exp[-i(\omega_{\sigma}-i\kappa/2)t]-\kappa\int_{0}^{t}ds\exp[-i(\omega_{\sigma}-i\kappa/2)(t-s)]\hat{\beta}_{{\rm in}}(s),
\end{equation}
and thus,
\begin{align}
\hat{\mathcal{B}}_{{\rm out}}(t)=\;&\hat{\beta}_{{\rm out}}(t)-\average{\hat{\beta}_{{\rm out}}(t)}\\
=\;&\hat{\mathcal{B}}_{{\rm in}}(t)+\sqrt{\kappa}\hat{\mathcal{B}}(0)\exp[-i(\omega_{\sigma}-i\kappa/2)t]-\kappa\int_{0}^{t}ds\exp[-i(\omega_{\sigma}-i\kappa/2)(t-s)]\hat{\mathcal{B}}_{{\rm in}}(s),
\end{align}
with $\hat{\mathcal{B}}(t)=\hat{\beta}(t)-\average{\hat{\beta}(t)}$ being the quantum fluctuation operator of the Bogoliubov mode $\hat{\beta}$. As can be verified by a straightforward calculation, the correlations for $\hat{\mathcal{B}}_{{\rm out}}(t)$ are:
\begin{gather}
\average{\hat{\mathcal{B}}_{{\rm out}}(t)\hat{\mathcal{B}}_{{\rm out}}^{\dagger}(t^{\prime})}=\delta(t-t^{\prime}),\\
\average{\hat{\mathcal{B}}_{{\rm out}}^{\dagger}(t)\hat{\mathcal{B}}_{{\rm out}}(t^{\prime})}=\average{\hat{\mathcal{B}}_{{\rm out}}(t)\hat{\mathcal{B}}_{{\rm out}}(t^{\prime})}=\average{\hat{\mathcal{B}}_{{\rm out}}^{\dagger}(t)\hat{\mathcal{B}}_{{\rm out}}^{\dagger}(t^{\prime})}=0,
\end{gather}
It then follows, by using the Bogoliubov transformation 
\begin{align}
\hat{\mathcal{A}}_{{\rm out}}(t)=\;&\hat{a}_{\rm out}(t)-\average{\hat{a}_{\rm out}(t)}\\
=\;&\cosh(r)\hat{\mathcal{B}}_{{\rm out}}(t)-e^{i\theta}\sinh(r)\hat{\mathcal{B}}_{{\rm out}}^{\dagger}(t),
\end{align}
that the correlations for $\hat{\mathcal{A}}_{{\rm out}}(t)$ are given by
\begin{gather}
\average{\hat{\mathcal{A}}_{{\rm out}}^{\dagger}(t)\hat{\mathcal{A}}_{{\rm out}}(t^{\prime})}=\sinh^{2}(r)\delta(t-t^{\prime}),\quad \quad \average{\hat{\mathcal{A}}_{{\rm out}}(t)\hat{\mathcal{A}}_{{\rm out}}^{\dagger}(t^{\prime})}=\cosh^{2}(r)\delta(t-t^{\prime}),\\
\average{\hat{\mathcal{A}}_{{\rm out}}(t)\hat{\mathcal{A}}_{{\rm out}}(t^{\prime})} =-\frac{1}{2}e^{i\theta}\sinh(2r)\delta(t-t^{\prime}),\quad \quad \average{\hat{\mathcal{A}}_{{\rm out}}^{\dagger}(t)\hat{\mathcal{A}}_{{\rm out}}^{\dagger}(t^{\prime})}=-\frac{1}{2}e^{-i\theta}\sinh(2r)\delta(t-t^{\prime}).
\end{gather}
Here, we have assumed that at the initial measurement time $t=0$, the cavity field, subject to a two-photon driving and a squeezed reservoir, is already in a steady state, i.e., the vacuum state of the mode $\hat{\beta}$, such that $\average{\hat{\mathcal{B}}^{\dagger}(t_{0})\hat{\mathcal{B}}(t_{0})}=\average{\hat{\mathcal{B}}(t_{0})\hat{\mathcal{B}}(t_{0})}=0$. From Eq.~(\ref{eq-meaurement-noise}), the measurement noise $\average{\hat{M}_{N}^{2}}$ takes the simple form
\begin{equation}\label{eq-measurement-noise}
\average{\hat{M}_{N}^{2}}=\kappa\tau\left[\cosh(2r)-\cos(2\phi_{h}-\theta)\sinh(2r)\right].
\end{equation}
Clearly, for $2\phi_{h}-\theta=0$, we obtain
\begin{equation}\label{eq-measurement-noise-exponentially-suppressed}
\average{\hat{M}_{N}^{2}}=\kappa\tau\exp(-2r)=\average{\hat{M}_{N}^{2}}_{\rm std}\exp(-2r),
\end{equation}
indicating the measurement noise is exponentially suppressed at any measurement time. Here, $\average{\hat{M}_{N}^{2}}_{\rm std}=\kappa\tau$ is the measurement noise of the standard readout with no squeezing. This result is in sharp contrast to the case of using IES or ICS alone as discussed above.    

Having achieved an exponentially suppressed measurement noise, let us now consider the measurement signal. We find from Eq.~(\ref{eq-bogoliubov-transformation-beta-in}) that
\begin{equation}
\average{\hat{\beta}_{\rm in}(t)}=\alpha_{{\rm in}}\exp(i\phi_{{\rm in}})\left\{\cosh(r)+\sinh(r)\exp[-i(2\phi_{{\rm in}}-\theta)]\right\}.
\end{equation}
Here, we have assumed that $\average{\hat{a}_{\rm in}\left(t\right)}=\alpha_{\rm in}e^{i\phi_{\rm in}}$. Since the signal separation is proportional to $|\average{\hat{\beta}_{\rm in}(t)}|$ (see below), we thus choose $2\phi_{{\rm in}}-\theta=0$, so as to ensure an exponential increase of $|\average{\hat{\beta}_{\rm in}(t)}|$ with the squeezing parameter $r$, yielding
\begin{equation}
\average{\hat{\beta}_{\rm in}(t)}=\alpha_{{\rm in}}\exp(r)\exp(i\phi_{{\rm in}}).
\end{equation} 
Then, according to Eq.~(\ref{eq-output-field-beta}), we obtain 
\begin{equation}
\average{\hat{\beta}_{{\rm out}}(t)}=\alpha_{{\rm in}}\exp(r)\exp(i\phi_{{\rm in}})\left\{ 1+\frac{i\kappa}{\omega_{\sigma}-i\kappa/2}\left\{1-\exp[-i(\omega_{\sigma}-i\kappa/2)t]\right\}\right\},
\end{equation}
under the initial condition of $\average{\hat{\beta}(0)}=0$. Subsequently, the measurement signal, defined in Eq.~(\ref{eq-measurement-signal}), is found by taking the Bogoliubov transformation $\hat{a}_{{\rm out}}(t)=\cosh(r)\hat{\beta}_{{\rm out}}(t)-e^{i\theta}\sinh(r)\hat{\beta}_{{\rm out}}^{\dagger}(t)$:
\begin{align}\label{esq-signal-separation-enhance}
\average{\hat{M}}=\;&\frac{2\alpha_{{\rm in}}e^{r}}{\sqrt{\kappa}}\bigg\{\left[2-\kappa\tau+2\cos(2\psi_{\sigma})\right]\left[\cos(\vartheta_{+})\cosh(r)-\cos(\vartheta_{-}+\theta)\sinh(r)\right]\nonumber\\
\;&-4e^{-\kappa\tau/2}\cos^{2}(\psi_{\sigma})\left[\cos(\vartheta_{+}+\omega_{\sigma}\tau)\cosh(r)-\cos(\vartheta_{-}+\theta+\omega_{\sigma}\tau)\sinh(r)\right]\bigg\},
\end{align}
where $\vartheta_{\pm}=2\psi_{\sigma}\pm\phi_{h}-\phi_{{\rm in}}$.
We now divide the signal separation, $|\average{\hat{M}}_{\uparrow}-\average{\hat{M}}_{\downarrow}|$, into two components, one along the measurement direction of homodyne detection (i.e., the squeezing direction), labelled $|\average{\hat{M}}_{\uparrow}-\average{\hat{M}}_{\downarrow}|_{\parallel}$; and the other along the direction perpendicular to the measurement direction of homodyne detection (i.e., the antisqueezing direction), labelled $|\average{\hat{M}}_{\uparrow}-\average{\hat{M}}_{\downarrow}|_{\perp}$. It can be seen that $|\average{\hat{M}}_{\uparrow}-\average{\hat{M}}_{\downarrow}|_{\parallel}$ and $|\average{\hat{M}}_{\uparrow}-\average{\hat{M}}_{\downarrow}|_{\perp}$ are found by setting $2\phi_{h}-\theta=0$, $\phi_{{\rm in}}-\phi_{h}=0$ and $\theta-2\phi_{h}=\pi$, $\phi_{\rm in}-\phi_{h}=\frac{\pi}{2}$, respectively, yielding
\begin{align}
\label{eq-signal-separation-sq}
&|\average{\hat{M}}_{\uparrow}-\average{\hat{M}}_{\downarrow}|_{\parallel}\nonumber\\
&=\frac{2\alpha_{{\rm in}}}{\sqrt{\kappa}}\bigg|\left[2-\kappa\tau+2\cos(2\psi_{-1})+2\cos(2\psi_{+1})\right]\left[\cos(2\psi_{-1})-\cos(2\psi_{+1})\right]\nonumber\\
&\quad\;-e^{-\kappa\tau/2}\left[\cos(\omega_{-1}\tau)+2\cos(2\psi_{-1}+\omega_{-1}\tau)+\cos(4\psi_{-1}+\omega_{-1}\tau)-4\cos^{2}(\psi_{+1})\cos(2\psi_{+1}+\omega_{+1}\tau)\right]\bigg|,\\
\label{eq-signal-separation-antisq}
&|\average{\hat{M}}_{\uparrow}-\average{\hat{M}}_{\downarrow}|_{\perp}\nonumber\\
&=\frac{2\alpha_{{\rm in}}e^{2r}}{\sqrt{\kappa}}\bigg|\left[2-\kappa\tau+2\cos(2\psi_{-1})\right]\sin(2\psi_{-1})-\left[2-\kappa\tau+2\cos(2\psi_{+1})\right]\sin(2\psi_{+1})\nonumber\\
&\quad\;-e^{-\kappa\tau/2}\left[\sin(\omega_{-1}\tau)+2\sin(2\psi_{-1}+\omega_{-1}\tau)+\sin(4\psi_{-1}+\omega_{-1}\tau)-4\cos^{2}(\psi_{+1})\sin(2\psi_{+1}+\omega_{+1}\tau)\right]\bigg|,
\end{align}
respectively.

\begin{figure*}[t]
	\centering
	\includegraphics[width=13cm]{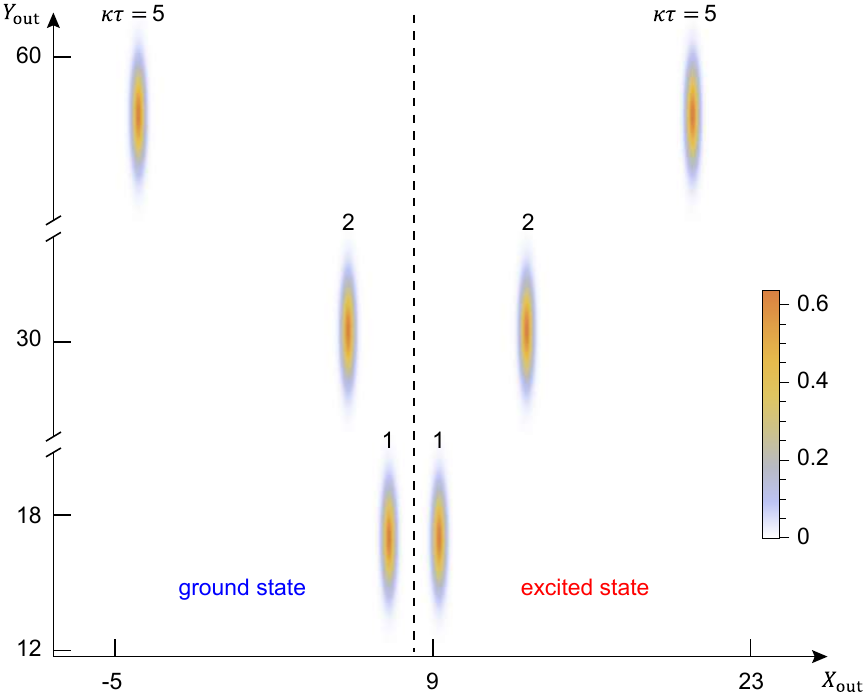}
	\caption{Phase-space representation of DQR simultaneously using IES and ICS. The Wigner functions on the left- and right-hand sides of the vertical dashed line correspond to the ground and excited states of the qubit, respectively. Here, we assumed $\chi=0.5\kappa$, $r=1$ and chose three different measurement times, i.e., $\kappa\tau=1$, $2$, $5$, as an example.}\label{fig_Wigner_function_IS_ES}
\end{figure*}
Intuitively, we can directly maximize $|\average{\hat{M}}_{\uparrow}-\average{\hat{M}}_{\downarrow}|_{\parallel}$ so as to maximize the SNR, but in this case, $|\average{\hat{M}}_{\uparrow}-\average{\hat{M}}_{\downarrow}|_{\perp}$, which is usually zero in the case of using IES or ICS, may be nonzero. For example, for a give measurement time $\kappa\tau=1$ and a given dispersive coupling $\chi=0.5\kappa$, the maximum value of $|\average{\hat{M}}_{\uparrow}-\average{\hat{M}}_{\downarrow}|_{\parallel}$ is $\simeq0.47\alpha_{\rm in}/\sqrt{\kappa}$ with $\tan(\psi_{+1})\simeq6.5$ and $\tan(\psi_{-1})\simeq4.5$; but at the same time, the value of $|\average{\hat{M}}_{\uparrow}-\average{\hat{M}}_{\downarrow}|_{\perp}$ is found to be $\simeq1.1\alpha_{\rm in}/\sqrt{\kappa}$. Thus for a fair comparison with the two cases of using IES and ICS separately, we require
\begin{equation}\label{seq-zero-antisq-direction}
	|\average{\hat{M}}_{\uparrow}-\average{\hat{M}}_{\downarrow}|_{\perp}=0.
\end{equation}
In fact, for a given measurement time, we can exactly ensure this requirement
with an appropriate effective cavity frequency $\omega_{{\rm sq}}$. The dependence of $\omega_{{\rm sq}}$ on $\kappa\tau$ is plotted in the inset in Fig.~2(a) in the main article. Furthermore, in Fig.~2(b) in the main article, we demonstrate an exponential enhancement in the SNR under the condition in Eq.~(\ref{seq-zero-antisq-direction}). This enhancement can be understood more deeply in the phase-space representation in Fig.~\ref{fig_Wigner_function_IS_ES}, which is in sharp contrast to the separate uses of IES and ICS shown in Figs.~\ref{fig_Wigner_function_ES}(c) and~\ref{fig_Wigner_function_IS}(c). 

We consider below the SNR in the two limits of $\kappa\tau\rightarrow0$ and $\infty$. In the limit of $\kappa\tau\rightarrow0$, the effective cavity frequency $\omega_{{\rm sq}}$ can be found from the condition in Eq.~(\ref{seq-zero-antisq-direction}),
\begin{equation}
\omega_{{\rm sq}}\simeq\frac{2.58}{\tau},
\end{equation}
which is inversely proportional to the measurement time $\tau$ [see inset in Fig.~2(a) in the main article]. As a consequence, we have 
\begin{equation}\label{seq-signal-separation-short-time}
|\average{\hat{M}}_{\uparrow}-\average{\hat{M}}_{\downarrow}|=|\average{\hat{M}}_{\uparrow}-\average{\hat{M}}_{\downarrow}|_{\parallel}\simeq\frac{0.27\alpha_{\rm in}}{\sqrt{\kappa}}\tan(\psi_{\rm sq})\left(\kappa\tau\right)^{3}\simeq0.81\exp(r)|\average{\hat{M}}_{\uparrow}-\average{\hat{M}}_{\downarrow}|_{\rm std},
\end{equation}
where $\tan(\psi_{\rm sq})=2\chi_{\rm sq}/\kappa$, such that the SNR in the limit of $\kappa\tau\rightarrow0$ is given by
\begin{equation}\label{seq-SNR-short-time}
{\rm SNR}\simeq0.81\exp(2r){\rm SNR}_{\rm std}.
\end{equation}
Here, $|\average{\hat{M}}_{\uparrow}-\average{\hat{M}}_{\downarrow}|_{\rm std}$ and ${\rm SNR}_{{\rm std}}$ refer to the signal separation and the SNR, respectively, of the standard readout with no squeezing. It can be surprisingly seen from Eq.~(\ref{seq-SNR-short-time}) that compared to the standard readout, the SNR can be {\it exponentially improved with $2r$, rather than $r$} as usually expected. Such a giant improvement arises from two contributions. The first contribution comes from the exponentially suppressed measurement noise as in Eq.~(\ref{eq-measurement-noise-exponentially-suppressed}), and the second one is due to the exponentially amplified dispersive coupling $\chi_{\rm sq}$ as in Eq.~(7) in the main article and thus the exponentially amplified signal separation as in Eq.~(\ref{seq-signal-separation-short-time}).

Furthermore, in the limit of $\kappa\tau\rightarrow\infty$, the condition in Eq.~(\ref{seq-zero-antisq-direction}) gives
\begin{equation}
\omega_{{\rm sq}}\simeq\frac{\kappa}{2}\sec(\psi_{\rm sq}),
\end{equation}
which is independent of the measurement time [see inset in Fig.~2(a) in the main article]. This yields
\begin{equation}\label{seq-signal-separation-long-time}
	|\average{\hat{M}}_{\uparrow}-\average{\hat{M}}_{\downarrow}|=|\average{\hat{M}}_{\uparrow}-\average{\hat{M}}_{\downarrow}|_{\parallel}\simeq\frac{4\alpha_{\rm in}}{\sqrt{\kappa}}\sin(\psi_{\rm sq})\kappa\tau\simeq\frac{\sin(\psi_{\rm sq})}{\sin(2\psi)}|\average{\hat{M}}_{\uparrow}-\average{\hat{M}}_{\downarrow}|_{\rm std},
\end{equation}
and then
\begin{equation}\label{seq-SNR-long-time}
	{\rm SNR}\simeq\frac{\sin(\psi_{\rm sq})}{\sin(2\psi)}\exp(r){\rm SNR}_{\rm std}.
\end{equation}
Equation~(\ref{seq-SNR-long-time}) indicates that in the limit $\kappa\tau\rightarrow\infty$, the SNR can have an exponential improvement with the squeezing parameter $r$. Note that the signal separation in Eq.~(\ref{seq-signal-separation-long-time}) is not significantly changed with increasing $r$, compared to the standard readout.
This is in contrast to the case of $\kappa\tau\rightarrow0$. Thus, along with an exponentially suppressed measurement noise given in Eq.~(\ref{eq-measurement-noise-exponentially-suppressed}), the SNR in the limit $\kappa\tau\rightarrow\infty$ can be improved exponentially with $r$, instead of $2r$, as seen in Eq.~(\ref{seq-SNR-long-time}). For typical parameters $e^{r}=10$ and $\chi=\kappa/2$, we can obtain $\sin(\psi_{\rm sq})\simeq\sin(2\psi)$ and, thus, ${\rm SNR}\simeq\exp(r){\rm SNR}_{\rm std}$ in the limit $\kappa\tau\rightarrow\infty$.

Hence, according to the above discussions, we see that with increasing the measurement time from $\kappa\tau\rightarrow0$ to $\kappa\tau\rightarrow\infty$, the improvement of the SNR is gradually changed from $\sim e^{2r}$ to $\sim e^{r}$, as shown in Fig.~2(b) in the main article. Furthermore, we see that the SNR improvement originates from two aspects, one of which is due to the measurement noise exponentially suppressed at any measurement time. The other aspect is due to the exponentially enhanced dispersive coupling, which can lead to an exponentially increased signal separation and thus SNR for short-time measurements, but which has almost no contribution to the improvement of the SNR for long-time measurements.

\section{Effects of parameter mismatches on the readout}
\label{Effects of parameter mismatches on the readout}
Our present proposal relies on the simultaneous use of IES and ICS, and further requires to satisfy the parameter conditions in Eq.~(\ref{seq-parameter-condition}). However, in realistic experiments, there are always some parameter mismatches, such that the conditions in Eq.~(\ref{seq-parameter-condition}) are not satisfied perfectly. In such an imperfect case, we assume that 
\begin{equation}\label{eq-parameter-imperfection}
	r_c=r+\delta_{r}, \quad {\rm and} \quad \theta-\varphi=\pi+\delta_{p},
\end{equation}
where $\delta_{r}$ and $\delta_{p}$ are the squeezing degree and direction mismatches, respectively. Below, we analyze the effects of these parameter mismatches on our readout proposal.

The ideal conditions in Eq.~(\ref{seq-parameter-condition}) lead to  
$\mathcal{N}=\mathcal{M}=0$, as mentioned in Sec.~\ref{Enhance dispersive qubit readout}; however, due to the parameter mismatches given in Eq.~(\ref{eq-parameter-imperfection}), $\mathcal{N}$ and $\mathcal{M}$ are no longer zero. Under such parameter mismatches, the correlations for the output noise operator $\hat{\mathcal{B}}_{{\rm out}}(t)=\hat{\beta}_{{\rm out}}(t)-\average{\hat{\beta}_{{\rm out}}(t)}$ are found to be
\begin{align}
	\left\langle\hat{\mathcal{B}}_{{\rm out}}(t)\hat{\mathcal{B}}_{{\rm out}}^{\dagger}(t^{\prime})\right\rangle =\;&	\left(\mathcal{N}+1\right)\delta(t-t^{\prime})-\kappa\mathcal{N}\exp\left[-\frac{1}{2}(\kappa_{\sigma}t+\kappa_{\sigma}^{*}t^{\prime})\right],\\
	\left\langle \hat{\mathcal{B}}_{{\rm out}}(t)\hat{\mathcal{B}}_{{\rm out}}(t^{\prime})\right\rangle =\;&	\mathcal{M}\delta(t-t^{\prime})-\frac{\kappa}{\kappa_{\sigma}}\mathcal{M}\exp\left[-\frac{1}{2}(t+t^{\prime})\kappa_{\sigma}\right]\begin{cases}
		(\kappa+i2\omega_{\sigma}e^{\kappa_{\sigma}t^{\prime}}) & {\rm if} \;\; t\geq t^{\prime},\\
		(\kappa+i2\omega_{\sigma}e^{\kappa_{\sigma}t}) & {\rm if} \;\;  t<t^{\prime},
	\end{cases}
\end{align}
where $\kappa_{\sigma}=\kappa+i2\omega_{\sigma}$.
It then follows that the measurement noise is given by
\begin{equation}
	\left\langle \hat{M}_{N}^{2}\right\rangle=\mathcal{R}_{0}+\mathcal{R}_{1}+\mathcal{R}_{2},
\end{equation}
where
\begin{align}
	\mathcal{R}_{0}=\;&\kappa\tau\left[\cosh(2r)+\cos(\varphi-\theta)\sinh(2r)\right],\\
	\mathcal{R}_{1}	=\;&8e^{-2r_{c}-\kappa\tau/2}\cos^2(\psi_{\sigma})\left[\cosh(\frac{\kappa\tau}{2})-\cos(\omega_{\sigma}\tau)\right]\left[1-\cosh(2r_{c})\cosh(2r)-\cosh(\theta-\varphi)\sinh(2r_{c})\sinh(2r)\right],\\
	\mathcal{R}_{2}=\;&e^{-2r_{c}-\kappa\tau}\sinh(2r_{0})\cos(\psi_{\sigma})\bigg\{e^{\kappa\tau}[(1-2\kappa\tau)\cos(\vartheta_1)-2(1-\kappa\tau)\cos(\vartheta_3)-3\cos(\vartheta_5)]\nonumber\\
	&+8e^{\kappa\tau/2}\cos(\psi_{\sigma})\cos(\vartheta_4+\omega_{\sigma}\tau)-4\cos^{2}(\psi_{\sigma})\cos(\vartheta_3+2\omega_{\sigma}\tau)\bigg\}.
\end{align}
Here, $\vartheta_n=n\psi_{\sigma}+\theta-\phi_{0}$. Moreover, we have set $2\phi_h-\theta=0$, and for compact notation, defined 
\begin{equation}
	\mathcal{N}=\sinh^2(r_0), \quad {\rm and} \quad \mathcal{M}=\frac{1}{2}e^{i\phi_{0}}\sinh(2r_{0}).
\end{equation}
Furthermore, the measurement signal $\average{\hat{M}}$ is the same as in the ideal case where $\delta_{p}=\delta_{r}=0$ [see Eq.~(\ref{esq-signal-separation-enhance})], but with a replacement $r\rightarrow r_{c}$.

\begin{figure}[t]
	\centering
	\includegraphics[width=7cm]{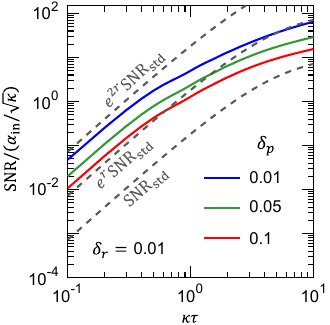}
	\caption{SNR in the presence of parameter mismatches as a function of the measurement time $\kappa\tau$ for $\delta_{p}=0.1$, $0.05$, $0.01$, and for $\delta_{r}=0.01$. Other parameters and what the three dashed curves represent are the same as in Fig.~2(b) in the main article.}\label{fig_parameter_imperfect}
\end{figure}

Having obtained the measurement noise and signal, we perform numerical simulations and plot in Fig.~\ref{fig_parameter_imperfect} the SNR in the presence of these parameter mismatches. It is seen that the exponential improvement in the SNR can still be achieved even for finite parameter mismatches, suggesting that our readout proposal is experimentally feasible.


%